\documentclass[journal=nalefd, manuscript=letter]{achemso}
\usepackage[version=3]{mhchem}
\usepackage{graphicx}
\usepackage{subfig}
\usepackage{amsmath}
\usepackage{color}
\usepackage{caption}
\usepackage{multirow}

\usepackage{graphics}
\usepackage{graphics}
\usepackage{amssymb}
\usepackage{amsfonts}
\usepackage{dcolumn}
\usepackage{dsfont}
\usepackage{latexsym}
\usepackage{rotating}
\usepackage{array}
\usepackage{makecell}
\usepackage{latexsym}
\usepackage{bbm}
\usepackage{physics}
\usepackage{float}
\usepackage{epsfig}
\usepackage{epsf}
\usepackage{psfrag}
\usepackage{bm}
\usepackage{amsthm}
\usepackage{eucal}
\usepackage{mathrsfs}
\usepackage{tabularx}
\usepackage{url}
\usepackage{braket}
\usepackage{epstopdf}
\usepackage[normalem]{ulem}
\usepackage{soul}
 
\author{Taylan Gorkan}

\affiliation{Materials Science and Engineering, School of Engineering for Matter, Transport, and Energy, Arizona State University}
\alsoaffiliation{Department of Physics, Adnan Menderes University, 09100 Aydin, Turkey}
\author{Jyotirish Das}
\affiliation{Department of Physics, Arizona State University, Tempe, AZ 85287, USA}

\author{Jesse Kapeghian}
\affiliation{Department of Physics, Arizona State University, Tempe, AZ 85287, USA}

\author{Muhammad Akram}
\affiliation{Department of Physics, Arizona State University, Tempe, AZ 85287, USA}
\alsoaffiliation{Department of Physics, Balochistan University of Information Technology, Engineering and Management Sciences (BUITEMS), Quetta 87300, Pakistan}

\author{Johannes V. Barth}
\affiliation{Physik Department E20, Technische Universit\"{a}t M\"{u}nchen, 85748 Garching, Germany}

\author{Sefaattin Tongay}
\affiliation{Materials Science and Engineering, School of Engineering for Matter, Transport, and Energy, Arizona State University}

\author{Ethem Akturk}
\affiliation{Department of Physics, Adnan Menderes University, 09100 Aydin, Turkey}
\alsoaffiliation{Physik Department E20, Technische Universit\"{a}t M\"{u}nchen, 85748 Garching, Germany}
\author{Onur Erten}
\affiliation{Department of Physics, Arizona State University, Tempe, AZ 85287, USA}
\email{onur.erten@asu.edu}

\author{Antia Botana}
\affiliation{Department of Physics, Arizona State University, Tempe, AZ 85287, USA}
\email{antia.botana@asu.edu}

\title[]{Skyrmion formation in Ni-based Janus dihalide monolayers: Interplay between magnetic frustration and Dzyaloshinskii-Moriya interaction}
  
\begin{document}





\begin{abstract}
We present a comprehensive theory of the magnetic phases in monolayer nickel-based Janus dihalides (NiIBr, NiICl, NiBrCl) through a combination of first-principles calculations and atomistic simulations. Phonon band structure calculations and finite temperature molecular dynamics simulations show that Ni-dihalide Janus monolayers
are 
stable. The parameters of the interacting spin Hamiltonian extracted from \textit{ab initio} calculations show that nickel-dihalide Janus monolayers exhibit varying degrees of magnetic frustration together with a large Dzyaloshinskii-Moriya interaction  due to their inherent inversion symmetry breaking. The atomistic simulations reveal a delicate interplay between these two competing magnetic interactions in giving rise to skyrmion formation. 


\end{abstract}

\maketitle


Skyrmions are nanoscale vortex-like topologically protected spin textures that  can be controlled by electric and thermal currents. These characteristics make them promising to store information in future memory and logic devices \cite{tokura, fert,ever,bogdanov}.  
Skyrmions traditionally arise from
the Dzyaloshinskii–Moriya interaction (DMI), driven by
spin–orbit coupling (SOC) in systems lacking inversion
symmetry \cite{MnSi,Munzer_PRB2010,Yu_NatMat2011, yang, Dupe_NatComm2016, Banerjee_PRX2014}. 
An alternative route to stabilize skyrmions takes place in geometrically frustrated centrosymmetric lattices wherein competing exchange interactions (i.e. the combination of magnetic frustration and weak easy-axis anisotropy) can be exploited instead \cite{Wang_AdvMat2016, Lin_PRLL2018, Yu_PNAS2012, Khanh_NatNano2020}. 
In the context of two-dimensional (2D) van der Waals magnets, Amoroso \textit{et al.}\cite{amoroso2020spontaneous} recently reported that via this latter type of mechanism several chiral magnetic phases including antibiskyrmions (A2Sk) and skyrmions (SkX) can be spontaneously stabilized in monolayer NiI$_2$ with an underlying triangular lattice for its magnetic ions. The magnetic frustration arises in this case from competing ferromagnetic (FM) and antiferromagnetic (AF) interactions. 
Indeed, it has been recently shown that NiI$_2$ exhibits a helimagnetic ground state at low temperatures that  persists all the way from the bulk down to monolayer limit \cite{Song_Nature2022}. 
 
\begin{figure}
\centering
\includegraphics[width=15.0cm]{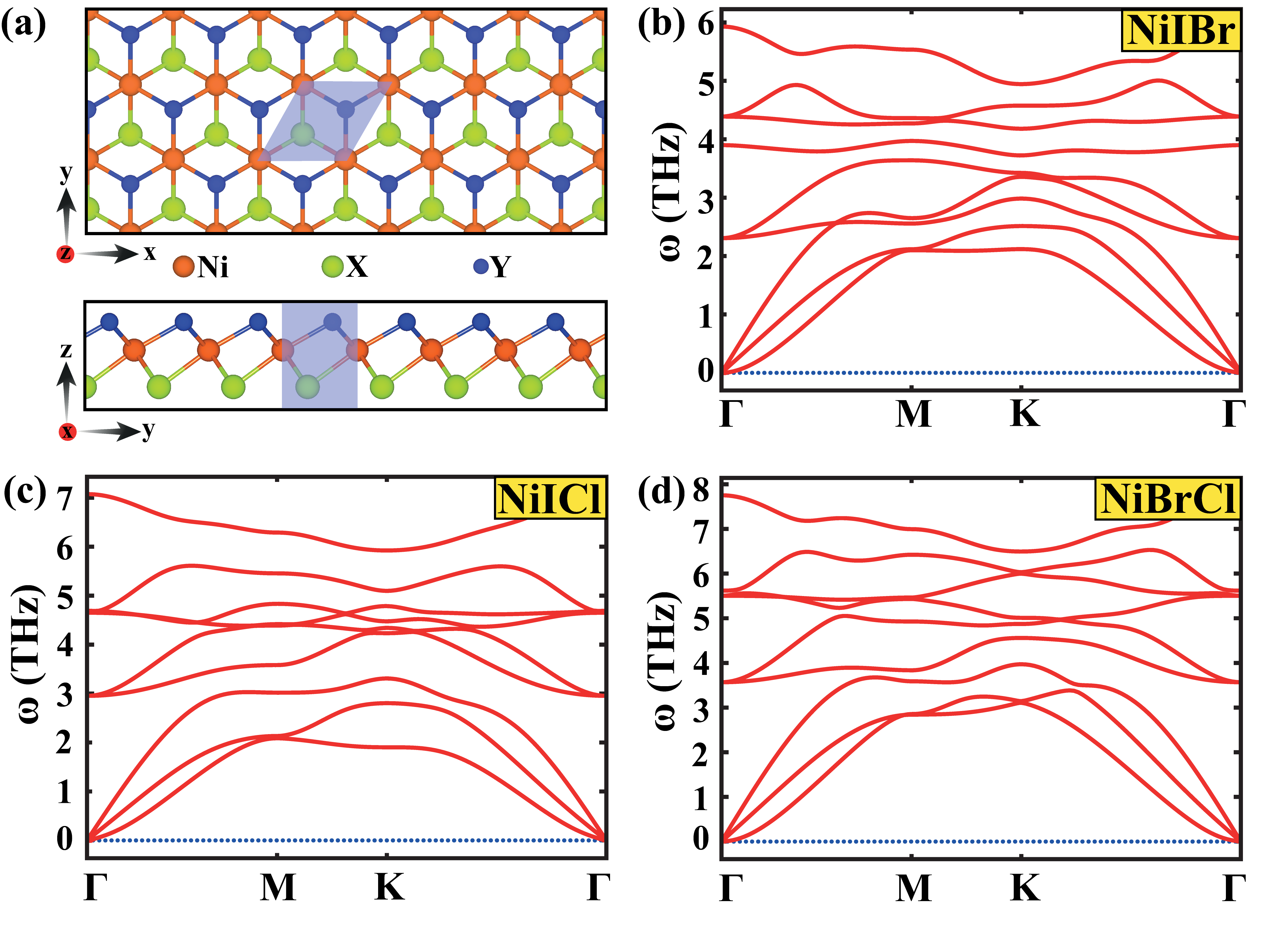}
\caption{(a) Top and side view of the optimized structure of Ni-based dihalide Janus monolayers. Orange spheres represent the Ni atoms, blue and green spheres represent the halide atoms (Cl, Br, I). The 2D primitive unit cell is shaded. (b), (c) and (d) Phonon dispersion curves along high-symmetry directions of the Brillouin zone calculated for NiIBr, NiICl, and NiBrCl.}
\label{fig:fig1}
\end{figure}

Even though generating skyrmions through DMI or magnetic frustration has its own merits, the interplay between the two mechanisms is not well-established. Here, using a combination of first-principles calculations and atomistic simulations we investigate this interplay on the chiral magnetic phases of nickel-based Janus monolayers NiXY (X/Y= Cl, Br, I, X $\neq$ Y). In Janus compounds, inversion symmetry is inherently broken\cite{Qin_AdvMat2022,Liang_PRB2020,Xu_PRB2020} as different halide anions occupy the top and bottom layers of Ni atoms as shown in Fig. \ref{fig:fig1} (a). This inherent inversion symmetry breaking has been shown to lead to a built-in polarization and a significant DMI interaction in several Janus monolayers\cite{Qin_AdvMat2022,Liang_PRB2020,Xu_PRB2020,Yuan_PRB2020,zhang2020emergence}.
By combining first-principles calculations and atomistic simulations, we find that in Ni-based Janus dihalides noncollinear magnetic states,
including skyrmion crystals, can be obtained  from the interplay between magnetic frustration and the DMI. 

\section{Results and Discussion}
\textbf{Structural properties and stability of NiXY monolayers.}
We start by analyzing the dynamical and thermal stability of NiXY monolayers (X being the heavier halide ion). The first-principles simulations of the
phonon spectrum indicate that NiICl, NiIBr, and NiBrCl  monolayers are dynamically stable, as shown in Fig. \ref{fig:fig1}. Using finite temperature \textit{ab-initio} molecular dynamics calculations (see Supporting Information) the thermal stability of these monolayers is also revealed. The relaxed structural parameters of
NiXY monolayers including lattice constants ($a$) and bond lengths (d$_{X(Y)}$) are shown in Table \ref{table1} in contrast to those of NiI$_2$. The lattice constants of
NiXY decrease as a function of the sum of X and Y ionic radii, and for a given NiXY
monolayer, d$_X$ is always larger than d$_Y$, due to the larger atomic radius of the X ion, as expected. The asymmetry between the top and bottom layers breaks the inversion symmetry, as mentioned above, thus allowing the DMI to arise. Further details on the calculations can be found in the Methods section.  

\begin{table}[]
\begin{tabular}{llll}
       & $a$(\AA)      & d$_X$(\AA)   & d$_Y$(\AA)   
       \\ \hline
NiI$_2$   & 3.96 & 2.74 & 2.74   
\\ \hline
NiIBr  & 3.84 & 2.71 & 2.60 
\\ \hline
NiICl  & 3.75 & 2.69 & 2.49 
\\ \hline
NiBrCl & 3.59 & 2.53 & 2.43 
\\ \hline         
\end{tabular}
\captionof{table}{Optimized in-plane lattice constant ($a$) and bond lengths of NiXY monolayers after the first-principles structural relaxation (d$_X$ from Ni to the heavier halide ion and d$_Y$ from Ni to the lighter halide ion). 
}
\label{table1}
\end{table}

\noindent
\textbf{Microscopic magnetic model.} The magnetic interactions between localized spins (\textbf{s$_i$}) in Ni dihalide monolayers can be modeled by the following spin Hamiltonian: 

\begin{equation}
    H= \frac{1}{2} \sum_{i \neq j} \mathbf{s_i \cdot J_{ij} \cdot s_j} + \frac{1}{2} \sum_{i \neq j} \mathbf{D_{ij} \cdot (s_i \times s_j)} + \sum_{i} \mathbf{s_i \cdot A_{i} \cdot s_i}
    \label{lt}
\end{equation}

Here, \textbf{J}$_{ij}$ denotes the tensor for the exchange coupling interactions (that can be  decomposed into an isotropic coupling term and an anisotropic one, the latter also referred to as the two-site anisotropy), \textbf{D}$_{ij}$ corresponds to the DMI interaction, and \textbf{A}$_{i}$ denotes the single-ion anisotropy. Both the DMI and two-site anisotropy arise from the spin–orbit
coupling, with the former favoring the canting of spin pairs, and the latter tending to orient them along a specific direction. The factors of 1/2 in front of the first two terms in Eq. \ref{lt} are to compensate for double-counting.

In Table \ref{table2} we report these magnetic interactions for NiXY monolayers (as obtained by density-functional theory (DFT)-based calculations including an on-site Coulomb repulsion $U_{eff}$= 1 eV) via a four-state method, a procedure analogous to that implemented in Ref. \citenum{amoroso2020spontaneous} for NiI$_2$ (further details on the derivation of the magnetic parameters  are presented in the Methods and in the Supporting Information). Note that Table \ref{table2} shows the isotropic J$^{\rm n,iso}$ interactions (n=1,2,3 for first, second and third nearest-neighbor (NN) Ni pairs, respectively), obtained from the trace of the \textbf{J}$_{ij}$ matrix. J$^{\rm s}_{ij}$ represents the  two-site anisotropy terms  (J$^{\rm s}_{ij}$= (J$_{ij}$ + J$^{\rm T}_{ij}$)/2 – J$^{\rm iso}_{ij}$I, where I is a unit matrix), and are derived from the J$_{ij}$ terms. Except for the J$^{\rm s}_{\rm yz}$ and J$^{\rm s}_{\rm zy}$ terms, all other off-diagonal terms are nominally zero by symmetry. We also show the NiI$_2$ results as a reference, for which we obtain values that are very similar to those reported in Ref. \citenum{amoroso2020spontaneous} : a dominant ferromagnetic nearest-neighbor exchange interaction (J$^{\rm 1iso}$) is derived, with a comparable antiferromagnetic third nearest-neighbor exchange (J$^{\rm 3iso}$) and a J$^{\rm 3iso}$/ J$^{\rm 1iso}$ ratio $\sim$ -0.8.   J$_{\rm yz}$/ J$^{\rm 1iso}$ is also a relevant ratio to consider, as it measures the exchange anisotropy $\sim$ -0.2. We note a sign change is obtained for J$_{\rm yz}$ with respect to Ref. \citenum{amoroso2020spontaneous} but $\pm${J$_{\rm yz}$} or $\pm${J$_{\rm zy}$} affects only the helicity of skyrmions.
The main traces of these leading magnetic interactions in NiI$_2$ remain in the Janus monolayers, with a sizable drop in the values of the magnetic constants as the size of the halide ion (and as a consequence the effects of the spin-orbit coupling strength) is reduced  when going from Cl to I. Specifically, NiIBr has magnetic parameters that are quite similar to those of NiI$_2$ with J$^{\rm 3iso}$/J$^{\rm 1iso}=-0.90$ and J$_{\rm yz}$/J$^{\rm 1iso}=-0.17$. For NiICl J$^{\rm 3iso}$/J$^{\rm 1iso}=-1.52$ and J$_{\rm yz}$/J$^{\rm 1iso}=-0.27$. For NiBrCl J$^{\rm 3iso}$/J$^{\rm 1iso}=-0.47$ and J$_{\rm yz}$/J$^{\rm 1iso}=-0.02$.

\begin{table}[]
\begin{tabular}{ccccc}
                                                                & NiI$_2$ & NiIBr & NiICl & NiBrCl \\ \hline
J$^{\rm 1iso}$                                                  & -7.53   & -6.12 & -4.04 & -5.87  \\
J$^{\rm 2iso}$                                                  & 0.02    & -0.01 & -0.33 & -0.08  \\
J$^{\rm 3iso}$                                                  & 6.15    & 5.48  & 6.14  & 2.18   \\ \hline
 J$^{s}_{\rm xx}$                                                   & -1.18   & -0.70 & -0.67 & -0.07  \\
J$^{s}_{\rm yy}$                                                    & 1.56    & 0.61  & 0.38  & 0.06   \\
J$^{s}_{\rm zz}$                                                    & -0.37   & 0.09  & 0.29  & 0.01   \\
J$^{s}_{\rm yz}$                                                    & 1.63    & 1.04  & 1.07  & 0.10   \\ \hline
$D_{\parallel}$                                                 & 0       & -0.72 & -0.53 & -0.04  \\
$D_{\perp}$                                                     & 0       & -0.70 & -1.07 & 0.02   \\ \hline
J$^{\rm 3iso}$/J$^{\rm 1iso}$                                   & -0.82   & -0.90 & -1.52 & -0.47  \\ \hline
J$_{\rm yz}$/J$^{\rm 1iso}$                                   & -0.22   & -0.17 & -0.27 & -0.02  \\ \hline
\end{tabular}
\captionof{table}{Isotropic first, second and third nearest neighbor interactions, two-site anisotropy, and D vector components for NiI$_2$ and NiXY monolayers. The J$_{\rm yz}$/J$^{\rm 1iso}$ and J$^{\rm 3iso}$/J$^{\rm 1iso}$ ratio is also shown. All values are expressed in meV.}
\label{table2}
\end{table}

For the single-ion (or magneto-crystalline) anisotropy (SIA), we considered only the diagonal terms (A$_{zz}$-A$_{xx}$). The SIA values are reported in the  Supporting Information and they are negligible with respect to the two main isotropic interactions and the
two-site anisotropy.  
Turning to the DMI, due to the presence of inversion symmetry in monolayer  NiI$_2$, the DMI is nominally zero.  As inversion symmetry is broken in Janus monolayers, a DMI naturally arises with the  D$_i$  vector components derived from the magnetic exchange matrix $D_{i}= (J_{jk} - J_{kj})/2$. In Table \ref{table2} the DMI components are shown for the different NiXY Janus monolayers. As expected, larger values ($\sim$ 0.7-1 meV) are obtained for the larger asymmetry between anions (that is, for NICl).  

\begin{figure}
\centering
\includegraphics[width=8.4cm]{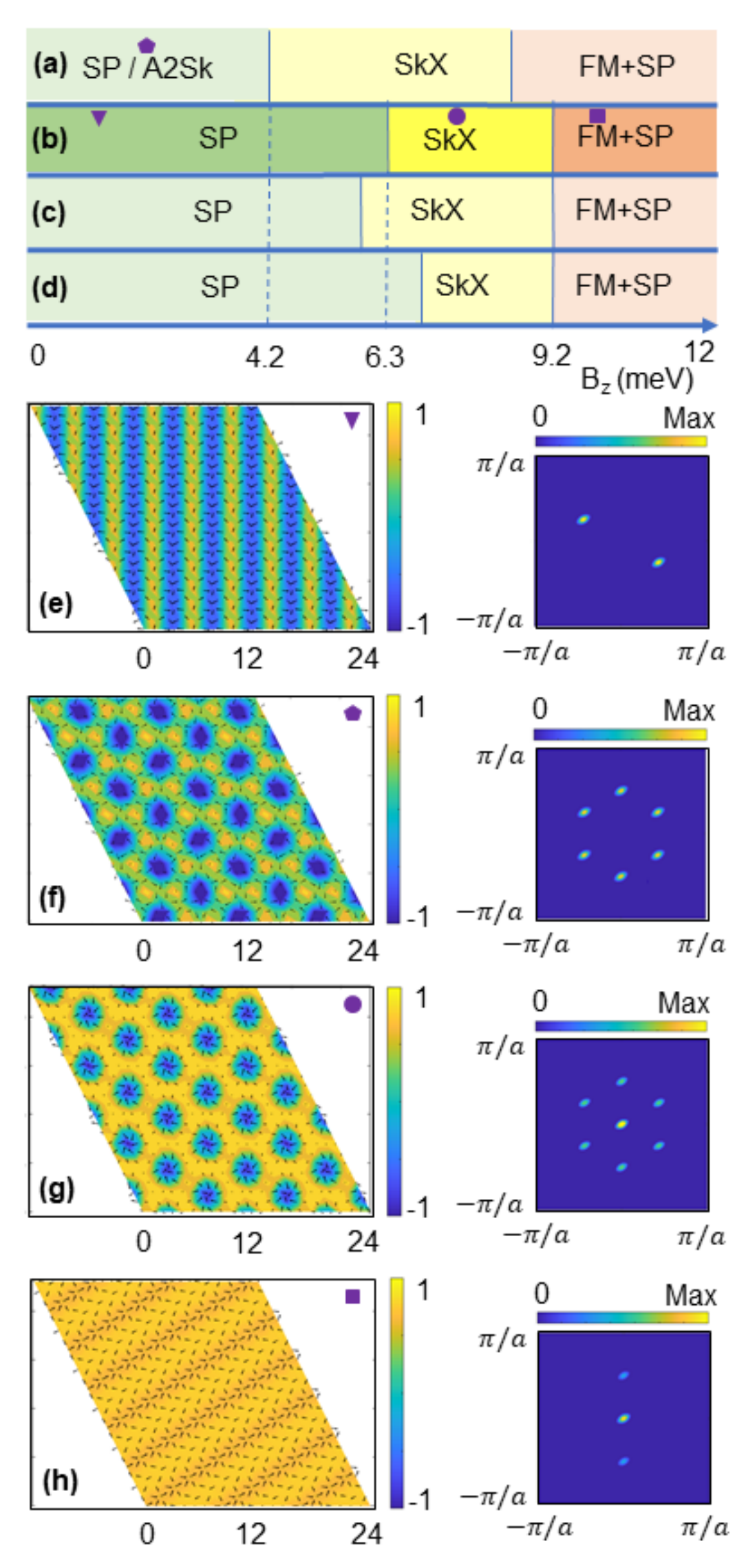}
\caption{Phase diagrams, magnetization textures and spin structure factors of NiIBr. The phase diagrams of NiIBr are presented as a function of $B$ for (a) $D_{\parallel}=0,~ D_{\perp}=0$, (b) $D=D^{\rm DFT}$, (c) $D_{\parallel}=D_{\parallel}^{\rm DFT}, D_{\perp}=0$ and (d) $D_{\parallel}=D_{\parallel}^{\rm DFT}, D_{\perp}=2D_{\perp}^{\rm DFT}$.
Magnetization texture and spin structure factor for (e) $D=D^{\rm DFT}$ and $B=1.2 \rm~ meV$, (f) $D=0$ and $B=2 \rm ~meV$, (g) $D=D^{DFT}$ and $B=7.3 \rm ~meV$, (h) $D=D^{DFT}$ and $B=9.8 ~\rm meV$.}
\label{fig_NiBrI_phasediagram}
\end{figure}

\begin{figure}
\centering
\includegraphics[width=8.4cm]{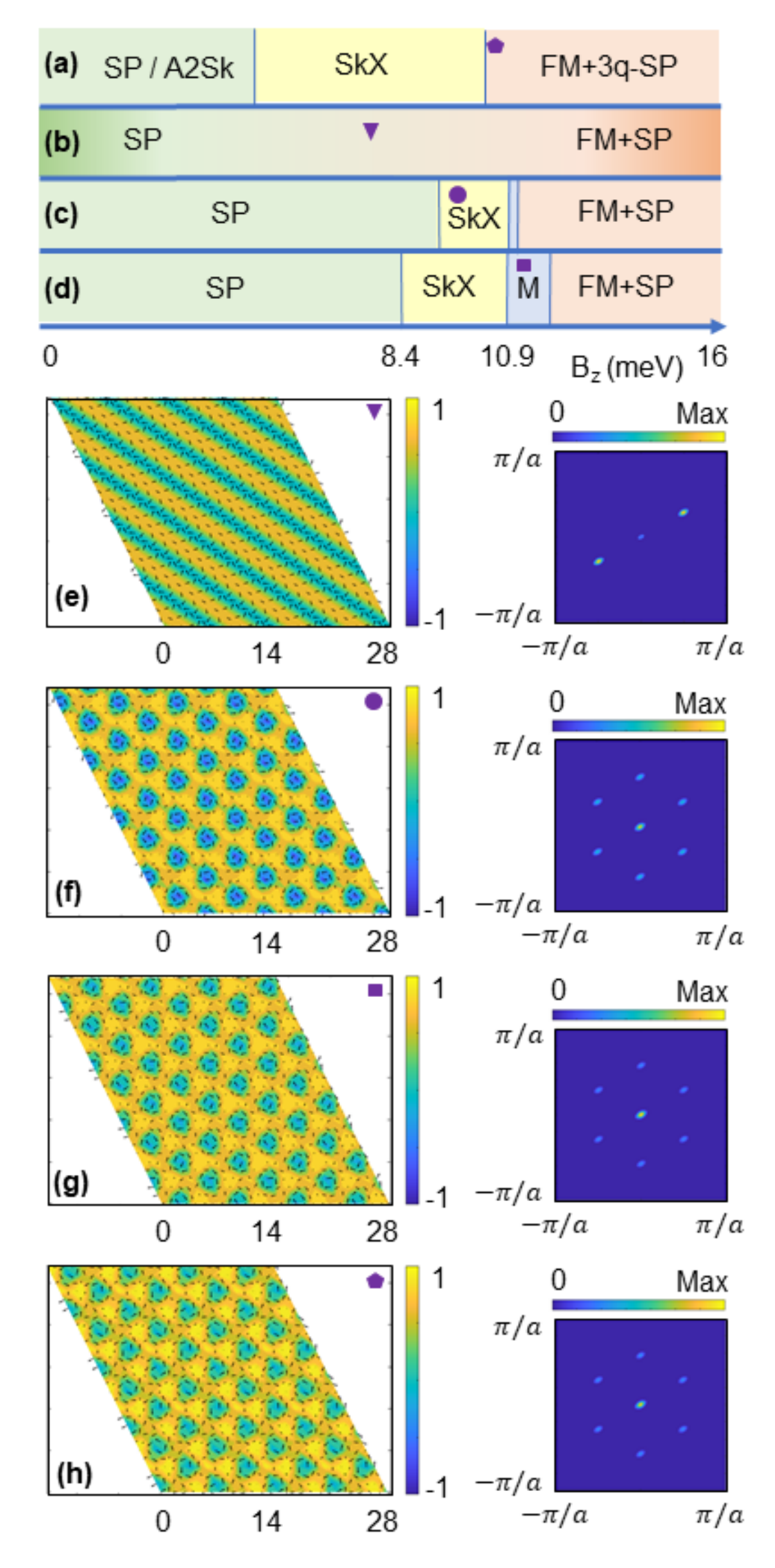}
\caption{Phase diagrams, magnetization textures, and spin structure factors of NiICl. The phase diagrams of NiICl are presented as a function of $B$ for (a) $D=0$, (b) $D = D^{DFT}$, (c) $D_{\parallel}=D_{\parallel}^{DFT}, D_{\perp}=D_{\perp}^{DFT}/2$ and (d) $D_{\parallel}=D_{\parallel}^{DFT}, D_{\perp}=0$.
Magnetization texture and spin structure factor for (e) $D=D^{DFT}$ and $B=7.7 \rm meV$, (f) $D_{\parallel}=D_{\parallel}^{DFT}, D_{\perp}=D_{\perp}^{DFT}/2$ and $B=9.8 \rm meV$, (g) $D_{\parallel}=D_{\parallel}^{DFT}, D_{\perp}=0$ and $B=11.2 \rm meV$, (h) $D=0$ and $B=10.5 \rm meV$.}
\label{fig_NiICl_phasediagram}
\end{figure}

\begin{figure}
\centering
\includegraphics[width=8.4cm]{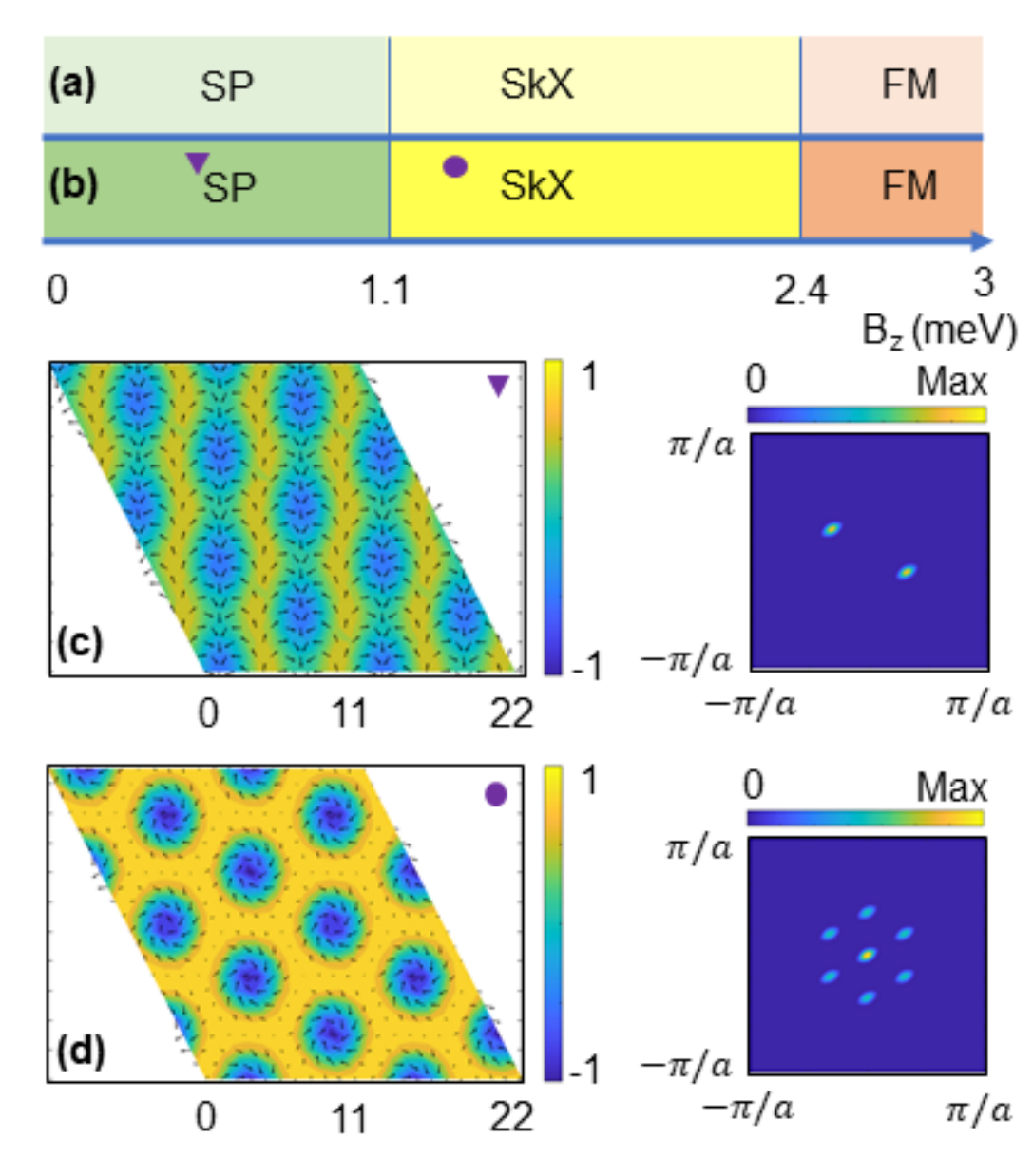}
\caption{Phase diagrams, magnetization textures, and spin structure factors of NiBrCl. The phase diagrams of NiBrCl are presented as a function of external magnetic field $B$ for (a) $D=0$, (b) $D=D^{DFT}$.
Magnetization texture and spin structure factor for (c) $D=D^{DFT}$ and $B=0.5 \rm meV$, (d) $D=D^{DFT}$ and $B=1.3 \rm meV$.}
\label{fig_NiBrCl_phasediagram}
\end{figure}

\noindent
{\bf Atomistic simulations.} Using the \textit{ab-initio} derived magnetic parameters, we solve the Landau-Lifshitz-Gilbert (LLG) equation\cite{1353448} in order to determine the T=0 phase diagram of Eq. \ref{lt}. Note that we tune the strength of the DMI (rather than using only the DFT-derived value) as experimentally such change can be achieved by using different substrates\cite{Zhang_PRB2021}. Further details of our calculations are presented in the Methods section. We obtain an approximate estimate of the size of the magnetic unit cell ($L\times L$) by the Luttinger-Tisza (LT) method and we verify these values  performing system-size dependent calculations. We find that the LT method provides a good estimate for the size of the magnetic unit cell $L$, and the DMI does not significantly change it. Our main results for the atomistic simulations are summarized in the phase diagrams of NiIBr, NiICl, and NiBrCl shown in Figs. \ref{fig_NiBrI_phasediagram}, \ref{fig_NiICl_phasediagram}, and \ref{fig_NiBrCl_phasediagram} respectively. Overall, we find a variety of magnetic phases including non-coplanar spirals (SP) for small and large magnetic field (B) and SkX for intermediate B. 

As mentioned above, NiI$_2$ exhibits a strong magnetic frustration due to the large $\mathrm{J^{3iso}}/\mathrm{J^{1iso}}$ as well as a sizable exchange anisotropy $\mathrm{J_{yz}}/\mathrm{J^{1iso}}$. Combining these two effects, Ref. \citenum{amoroso2020spontaneous} showed using Monte-Carlo calculations, that NiI$_2$ may show an antibiskyrmion crystal at low B and a skyrmion crystal at higher B. In the Supporting Information, we present our LLG-derived  phase diagram for NiI$_2$ which is consistent with that of Ref. \citenum{amoroso2020spontaneous}. 

In the following, in order to investigate the interplay between the magnetic frustration and DMI for the Janus materials, we first consider the case with $D=0$ (to compare with NiI$_2$) and include $D$ afterwards. We start our discussion of the atomistic simulations of Janus Ni-dihalides with the T= 0 phase diagram of NiIBr as a function of external magnetic field as shown in Fig. \ref{fig_NiBrI_phasediagram}. As mentioned above, for NiIBr, the exchange couplings obtained from DFT have similar values to those of NiI$_2$ \cite{amoroso2020spontaneous} with J$^{\rm 3iso}$/J$^{\rm 1iso}=-0.90$ and J$_{\rm yz}$/J$^{\rm 1iso}=-0.17$. Therefore, the phase diagram for $D=0$ is quite similar to that of NiI$_2$ \cite{amoroso2020spontaneous} with A2Sk and SP phases quasi-degenerate at low B (Fig. \ref{fig_NiBrI_phasediagram}(a)). We find that the SP has slightly lower energy by about $\sim 0.3$ meV, which is within the error bar of the DFT estimates. At $B =4.2$ meV, we obtain a SkX phase which survives up to $B=9$ meV. As the magnetic field is further increased, a FM+SP phase is derived with three peaks in the spin structure factor. This phase adiabatically connects with a fully-polarized FM one at larger fields. For finite DMI, the A2Sk phase completely vanishes and the domain of the SkX phase decreases as a function of DMI. To further investigate this effect, we consider three cases keeping the $D_\parallel$ equal to the value obtained by DFT ($D_{\parallel}=D_{\parallel}^{\rm DFT}$) while varying the $D_\perp = 2D_\perp^{\rm DFT}, D_\perp^{\rm DFT}$ and $D_\perp = 0$ as shown in Fig. \ref{fig_NiBrI_phasediagram}(b), (c) and (d). Note that $D_{\parallel}^{\rm DFT} \sim D_{\perp}^{\rm DFT}$ for NiIBr. From the phase diagrams, it is clear that $D_{\perp}$ suppresses the domain of stability of skyrmion formation. We find this trend holds for NiICl as well (see below). The underlying reason for this suppression stems on the fact that $D_{\perp}$ promotes spiral formation in the plane whereas SkX formation requires the spirals to propagate out of the plane. As a result, the presence of a $D_{\perp}$ term is in direct competition with SkX formation. The topological charge densities of all of the phases are presented in Supporting Information.

Next, we discuss the phase diagram of NiICl. In the absence of DMI (Fig. \ref{fig_NiICl_phasediagram}(a)), the phase diagram is similar to that of NiBrI with quasi-degenerate SP and A2Sk phases at low B, SkX at intermediate B, and a 3q spiral with a FM component (3q+FM) at large B. The 3q+FM state has topological charge zero but it has six peaks in the spin structure factor similar to the SkX phase. This state smoothly transitions to a FM one upon increasing the magnetic field, a behavior similar to that reported by Ref. \citenum{Amoroso_Nanomatt2021}. Once again, including the DMI diminishes the range of the A2Sk phase. Similar to NiIBr, we also find that $D_{\perp}$ suppresses SkX formation. As the $D_{\perp}$ value obtained from DFT calculations for NiICl is about twice as large as $D_{\parallel}$ ($D_{\perp}^{\rm DFT} \sim 2 D_{\parallel}^{\rm DFT}$) our atomistic simulations find that NiICl stays in a non-coplanar spiral for all external magnetic fields, completely skipping the intermediate SkX phase as shown in Fig. \ref{fig_NiICl_phasediagram}(b). We verify that the SkX reemerges for smaller $D_{\perp}$ as shown in Fig. \ref{fig_NiICl_phasediagram}(c) and (d) for $D_{\perp} = D_{\perp}^{\rm DFT}/2$ and $0$. We also note that for $D_{\perp} = D_{\perp}^{\rm DFT}/2$ and $D_{\perp}=0$, we observe a small region of a meron crystal phase. A meron is similar to half a skyrmion where the spin only wraps the upper or lower hemisphere, halving the topological charge\cite{Yu_Nat2018} (see Methods). 



Lastly, we discuss the phase diagram for NiBrCl presented in Fig. \ref{fig_NiBrCl_phasediagram}. Unlike the previous two Janus materials, NiBrCl has significantly smaller anisotropic symmetric  exchange $\mathrm{J_{yz}}/\mathrm{J^{1iso}} = -0.02$. Therefore, there is no A2Sk phase, even in the absence of DMI. A spiral is stabilized for small fields which evolves into a SkX phase and subsequently to a FM at higher fields. Moreover, the DMI is also significantly smaller, $D\sim0.04\rm ~ meV$, and we find that the $D$ values obtained from DFT have no impact on the phase diagram described above. 


\begin{figure}
    \centering
\includegraphics[width=\columnwidth]{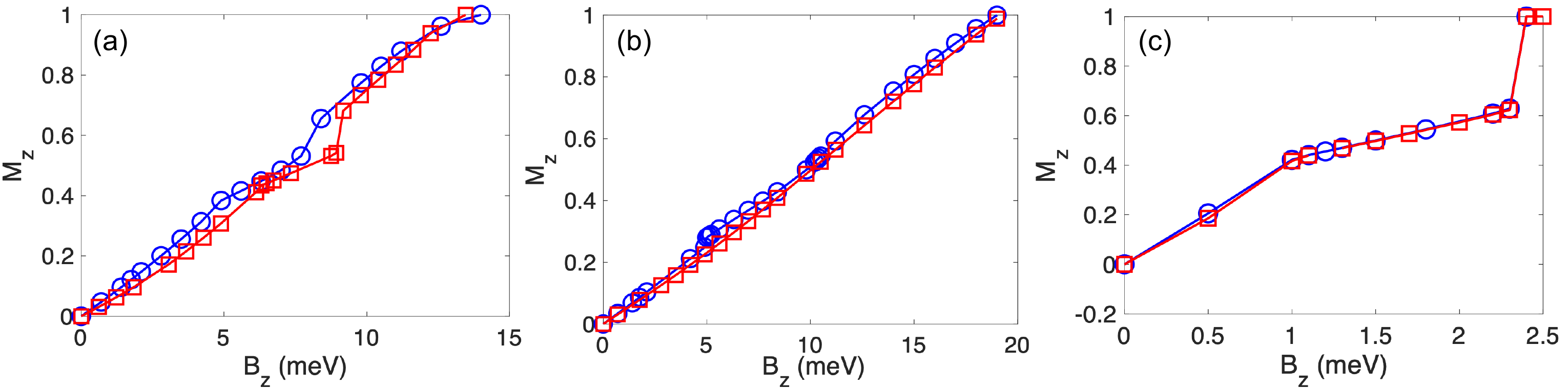}
    \caption{Normalized z-component of the average magnetization as a function of magnetic field for (a) NiIBr, (b) NiICl, (c) NiBrCl. The blue cirles and red squares correspond to $D=0$ and $D=D^{DFT}$ values respectively.}
    \label{m_vs_b}
\end{figure}

 
For 2D magnets, the $z$-component of the magnetization ($M_z$) can be directly measured via magnetic circular dichroism (MCD)\cite{Jiang_NatMatt2018} which can be used to deduce noncoplanar magnetic phases\cite{Xu_NatNano2022, Akram_NanoLett2021}. In Fig. \ref{m_vs_b}, we present $M_z$ for the three Janus materials as a function of external field with and without DMI. Sudden changes in $M_z$ denote a phase transition, and can be observed for NiIBr and NiBrCl. We do not find an abrupt change in $M_z$ for NiICl as this compound remains a spiral for all values of $B$. 

\section{Conclusions}
We have studied the magnetic phases of monolayer nickel-based Janus dihalides by a combination of first-principles calculations and atomistic simulations. We have shown that these monolayers are thermodynamically stable by using both phonon calculations and molecular dynamics simulations. We have extracted the magnetic exchange parameters from \textit{ab initio} calculations and used atomistic simulations to explore the magnetic phase diagram. We have found a variety of magnetic phases including non-coplanar spirals for small and large magnetic field and skyrmion crystal formation for intermediate B. Importantly, we have found that there is a fragile interplay between the DMI and magnetic frustration, particularly when it comes down to stabilizing a skyrmion phase. Interesting directions for future work include studying the effects of different stacking patterns and moir\'e superlattices of these Janus materials\cite{Akram_NanoLett2021, Akram_PRB2021}. Including the quantum effects arising from spin-1 local moments \cite{Zhang_arXiv2022} is  another interesting path to explore.

\section{Methods}
\subsection{Computational details of the density functional theory calculations and four-state method}
Density functional theory calculations were performed employing the Vienna Ab-initio Simulation Package, VASP\cite{vasp} using the   the Perdew-Burke-Ernzerhof (PBE)\cite{pbe} version of the generalized gradient approximation (GGA) with projected augmented wave (PAW)\cite{bloch, paw} potentials. We have considered $3d^{8}$ $4s^{2}$, $5s^{2}$ $5p^{5}$ ,  $4s^{2}$  $4p^{5}$ and $3s^{2}$  $3p^{5}$ configurations for Ni, I, Br and Cl, respectively, as valence electrons. The chosen energy cutoff of the plane-wave basis was 450 eV. Brillouin zone integration was performed using a 24$\times$24$\times$1 k-point grid. The energy convergence value was adjusted to be 10$^{-6}$ eV between two successive electronic steps. The vacuum spacing between Ni-based monolayers was taken to be 20 Å to avoid interactions between images.

By using the conjugate-gradient algorithm\cite{broyden1,broyden2}, all the atomic positions and lattice parameters were fully optimized until the total energy and forces were minimized and the maximum pressure in the unit cell was reduced to less than 0.5 kbar. All geometric relaxations were performed for a ferromagnetic state within PBE.

In order to compute the dynamical stability of the structures, we calculated force constants of the (4$\times$4$\times$1) supercell in real space using density functional perturbation theory (DFPT)\cite{dfpt} as implemented in PHONOPY\cite{phonopy}. These structures, having vibration modes with frequencies $w^{2}>0$ for all $k$ in the first BZ, were considered to be stable. 

We have also studied the thermal stability of Ni-based Janus dihalide monolayers by performing \textit{ab-initio} molecular dynamics (AIMD) calculations at finite temperatures. The time step between ionic iterations was taken to be $2 fs$. After every 50 steps, the velocities were rescaled to match the desired temperature. Further details on the AIMD calculations are shown in the Supporting Information.

 PBE+ $U$ (within the Dudarev's approach) \cite{dudarev} was used to account for the strong on-site interactions in the Ni-3$d$ electrons. To extract the first-principles-based magnetic parameters spin-orbit coupling (SOC) was also added. Different  $U$ values were used of 1,2 and 5 eV. In the main text we present the results of calculations performed at $U$= 1 eV (in agreement with Ref. \citenum{amoroso2020spontaneous}). The results for $U$= 2 and 5 eV are shown in the Supporting Information. All the derived magnetic moments per Ni atom are $\sim$1.5 ~$\mu_B$, consistent with the expected 2+ (high spin) oxidation state of Ni. Also, halide atoms I, Br and Cl develop magnetic moments of 0.27 $\mu_B$, 0.20 $\mu_B$, and 0.16 $\mu_B$, respectively.

To account for the magnetic interactions of the Ni pair, we use a four-state energy mapping method \cite{xiang,vsabani}. In this method, four different spin configurations are applied to a chosen Ni pair for all exchange coupling terms. 
The Hamiltonian (Eq. \ref{lt}) is solved to determine the exchange coupling tensor by using the PBE+SOC+U energies from the four-state energy mapping method. The DMI and two-site anisotropy terms can be derived from the exchange coupling terms. A  large enough supercell is chosen to avoid periodic interactions of Ni-pairs (for SIA, first and second nearest-neighbor interactions we used a 5$\times$4$\times$1 supercell and a 6$\times$3$\times$1 supercell was used for third nearest-neighbor interactions\cite{amoroso2020spontaneous}). Further details on the four-state mapping method are provided in the Supporting Information. Note that the exchange tensor we discuss in the main text is expressed in terms of the
cartesian \{x, y, z\} basis, where x is parallel to the
Ni-Ni bonding vector (see Supporting Information).
In order to better visualize the role of the two-site
anisotropy, it is useful to express the interaction within the local
principal-axes basis which diagonalize the exchange
tensor, as detailed in \citenum{amoroso2020spontaneous}, the corresponding eigenvalues are shown in the Supporting  Information.

\subsection{LLG Simulations}
To find the ground state of Hamiltonian (Eq. \ref{lt}), we solve the Landau-Lifshitz-Gilbert (LLG) equations: \cite{1353448}
\begin{eqnarray} 
\frac{d\textbf{s}}{dt}=-\gamma \textbf{s}\times \textbf{B}^{\rm eff}+\alpha \textbf{s} \times \frac{d\textbf{s}}{dt},
\label{llg}
\end{eqnarray}
where $\textbf{B}^{\rm eff}=-\delta H/\delta \textbf{s}$, $\gamma$ is the gyromagnetic ratio and $\alpha$ is Gilbert damping coefficient. We solve the LLG equations self-consistently keeping 
$|{\bf s}|=1$
and imposing periodic boundary conditions. 
The LLG equations were implemented through a semi-implicit midpoint method \cite{Mentink_2010} in MATLAB. We considered 200 random spin configurations as the initial state for a particular magnetic field and picked the lowest energy configuration after convergence. 

\subsection{Luttinger-Tisza method}
In order to get an estimate for the size of magnetic unit cell $(L \times L)$, we use the Luttinger-Tisza method\cite{Luttinger_PR1946} which replaces the hard spin constraint $|{\bf{s}}_i| = 1$ with a soft spin constraint $\sum_i |\mathbf{s_i}|^2 = N$. This allows us to find the lowest energy, coplanar spiral configurations with a wave vector $\bf q$ which sets a natural length scale in the problem $L\sim 2\pi/q$. This length scale is also important for SkX and A2Sk phases as they are a superposition of three spirals with the same $q$ but rotated by 120 degrees with respect to each other. 

For an isotropic model with only the first and third nearest neighbour interactions $\mathrm{J^{1iso}}$ and $\mathrm{J^{3iso}}$, it is possible to find an analytical expression for the wave vector $q=2 \cos^{-1}[(1+\sqrt{1-2 \mathrm{J^{1iso}}/\mathrm{J^{3iso}}})/4]$ \cite{PhysRevB.93.184413,Batista_2016}. Including the DMI, it is not possible to find an analytical expression for $\bf{q}$, therefore we numerically find the $\bf{q}$ that minimizes the energy. We obtain  
$L =~7.69~(7.89),~6.9~ (7.07), ~10.33~ (10.4)$ with (without) including DMI for NiIBr, NiICl and NiClBr, respectively. It is apparent that the DMI does not significantly change $L$. We take a multiple of $L$ that is integer and feasible for the LLG simulations. We have checked the validity of this method by varying the system size in the LLG simulations and we find that the Luttinger-Tisza method provides the correct $L$ in all cases.

\subsection{Topological charge and spin structure factor}
The topological charge of a continuous field $\mathbf{s}(x,y)$ is defined as
\begin{equation}
    Q=\frac{1}{4 \pi} \int d^2 \mathbf{r} \; \mathbf{s} \cdot \bigg( \frac{\partial \mathbf{s}}{\partial x} \times \frac{\partial \mathbf{s}}{\partial y} \bigg)
    \label{tQ_main}
\end{equation}
which signifies the number of times the spins wrap around a unit sphere \cite{nagaosa2013topological}. Using $\mathbf{s}=(\cos \Phi(\phi)\, \sin \Theta(r), \sin \Phi(\phi) \sin \Theta(r) , \cos \theta(r))$ and  $\mathbf{r}=(r \cos \phi, r \sin \phi)$,  Eq. \ref{tQ_main} simplifies to 
$Q=
  [\cos \Theta(r)] \Bigr|_{\theta (r=0)}^{\theta (r=\infty)} \, 
  [\Phi] \Bigr|_{\phi=0}^{\phi=2 \pi}
$ where the first part is called polarity and the second part is called vorticity\cite{Amoroso_Nanomatt2021}. For skyrmions  (biskyrmions),the spins wrap around once (twice) and therefore the vorticity (the $\phi$ term) is equal to one (two). On the other hand, for merons the angle between a spin at the center and at the edge is $90$ degrees compared to $180$ degrees for skyrmions. This halves the polarity (the cosine term) and therefore halves the topological charge. In order to evaluate the topological charge in the lattice model, we use the following definition\cite{BERG1981412} 
\begin{equation}
    \rm \tan \left( \frac{\Omega}{2} \right)=\frac{\mathbf{s}_{1} \cdot \mathbf{s}_{2} \times \mathbf{s}_{3}}{1+\mathbf{s}_{1} \cdot \mathbf{s}_{2}+\mathbf{s}_{2} \cdot \mathbf{s}_{3}+\mathbf{s}_{3} \cdot \mathbf{s}_{1}}
\end{equation}
where $\Omega$ is calculated over a triangular plaquette with spins $\mathbf{s}_{1}$, $\mathbf{s}_{2}$ and $\mathbf{s}_{3}$ and summed over the magnetic unit cell to give the topological charge $Q$.

Apart from the topological charge, we use the spin structure factor to determine the magnetic phases. The spin structure factor is defined as
\begin{equation}
S(\mathbf{q})=\frac{1}{N} \sum_{\alpha = x,y,z} \left\langle \left| \sum_{\mathrm{i}} s_{\mathrm{i},\alpha} e^{-i \mathbf{q} \cdot \mathbf{r}_{\mathrm{i}}} \right|^{2} \right\rangle ,
\end{equation}

where $N=L^{2}$ is the total number of spins and where the position of spin $s_{\mathrm{i}}$ is denoted by $\mathbf{r}_{\mathrm{i}}$.

\section{ASSOCIATED CONTENT}

\section{AUTHOR INFORMATION}
\textbf{Corresponding Authors:} Onur Erten and Antia Botana.

\noindent {\bf Author Contributions:}
Taylan Gorkan, Jyotirish Das, Jesse Kapeghian and Muhammad Akram contributed equally.

\textbf{Notes}
The authors declare no competing financial interest.

\section{Acknowledgements}
JD, JK, OE and AB acknowledge support from NSF Award
No. DMR 1904716. S.T acknowledges DMR-2206987 and DOE-SC0020653. MA is supported by Fulbright Scholarship. EA acknowledges the Alexander von Humboldt Foundation for a Research Fellowship for Experienced Researchers. TG was supported by the Scientific and Technological Research Council of Turkey (TUBITAK) under the 2214 scholarship.
Computational resources used in this work were provided by the TUBITAK
 (The Scientific and Technical Research Council of Turkey) ULAKBIM,
High Performance and Grid Computing Center (Tr-Grid e-Infrastructure), and the ASU
Research Computing Center for HPC resources.

\bibliography{Ni_references}

\providecommand{\latin}[1]{#1}
\makeatletter
\providecommand{\doi}
  {\begingroup\let\do\@makeother\dospecials
  \catcode`\{=1 \catcode`\}=2 \doi@aux}
\providecommand{\doi@aux}[1]{\endgroup\texttt{#1}}
\makeatother
\providecommand*\mcitethebibliography{\thebibliography}
\csname @ifundefined\endcsname{endmcitethebibliography}
  {\let\endmcitethebibliography\endthebibliography}{}
\begin{mcitethebibliography}{48}
\providecommand*\natexlab[1]{#1}
\providecommand*\mciteSetBstSublistMode[1]{}
\providecommand*\mciteSetBstMaxWidthForm[2]{}
\providecommand*\mciteBstWouldAddEndPuncttrue
  {\def\EndOfBibitem{\unskip.}}
\providecommand*\mciteBstWouldAddEndPunctfalse
  {\let\EndOfBibitem\relax}
\providecommand*\mciteSetBstMidEndSepPunct[3]{}
\providecommand*\mciteSetBstSublistLabelBeginEnd[3]{}
\providecommand*\EndOfBibitem{}
\mciteSetBstSublistMode{f}
\mciteSetBstMaxWidthForm{subitem}{(\alph{mcitesubitemcount})}
\mciteSetBstSublistLabelBeginEnd
  {\mcitemaxwidthsubitemform\space}
  {\relax}
  {\relax}

\bibitem[Tokura and Kanazawa(2020)Tokura, and Kanazawa]{tokura}
Tokura,~Y.; Kanazawa,~N. Magnetic skyrmion materials. \emph{Chemical Reviews}
  \textbf{2020}, \emph{121}, 2857--2897\relax
\mciteBstWouldAddEndPuncttrue
\mciteSetBstMidEndSepPunct{\mcitedefaultmidpunct}
{\mcitedefaultendpunct}{\mcitedefaultseppunct}\relax
\EndOfBibitem
\bibitem[Fert \latin{et~al.}(2017)Fert, Reyren, and Cros]{fert}
Fert,~A.; Reyren,~N.; Cros,~V. Magnetic skyrmions: advances in physics and
  potential applications. \emph{Nature Reviews Materials} \textbf{2017},
  \emph{2}, 1--15\relax
\mciteBstWouldAddEndPuncttrue
\mciteSetBstMidEndSepPunct{\mcitedefaultmidpunct}
{\mcitedefaultendpunct}{\mcitedefaultseppunct}\relax
\EndOfBibitem
\bibitem[Everschor-Sitte \latin{et~al.}(2018)Everschor-Sitte, Masell, Reeve,
  and Kl{\"a}ui]{ever}
Everschor-Sitte,~K.; Masell,~J.; Reeve,~R.~M.; Kl{\"a}ui,~M. Perspective:
  Magnetic skyrmions—Overview of recent progress in an active research field.
  \emph{Journal of Applied Physics} \textbf{2018}, \emph{124}, 240901\relax
\mciteBstWouldAddEndPuncttrue
\mciteSetBstMidEndSepPunct{\mcitedefaultmidpunct}
{\mcitedefaultendpunct}{\mcitedefaultseppunct}\relax
\EndOfBibitem
\bibitem[Bogdanov and Panagopoulos(2020)Bogdanov, and Panagopoulos]{bogdanov}
Bogdanov,~A.~N.; Panagopoulos,~C. Physical foundations and basic properties of
  magnetic skyrmions. \emph{Nature Reviews Physics} \textbf{2020}, \emph{2},
  492--498\relax
\mciteBstWouldAddEndPuncttrue
\mciteSetBstMidEndSepPunct{\mcitedefaultmidpunct}
{\mcitedefaultendpunct}{\mcitedefaultseppunct}\relax
\EndOfBibitem
\bibitem[M\"uhlbauer \latin{et~al.}(2009)M\"uhlbauer, Binz, Jonietz,
  Pfleiderer, Rosch, Neubauer, Georgii, and B\"oni]{MnSi}
M\"uhlbauer,~S.; Binz,~B.; Jonietz,~F.; Pfleiderer,~C.; Rosch,~A.;
  Neubauer,~A.; Georgii,~R.; B\"oni,~P. Skyrmion lattice in a chiral magnet.
  \emph{Science} \textbf{2009}, \emph{323}, 915--919\relax
\mciteBstWouldAddEndPuncttrue
\mciteSetBstMidEndSepPunct{\mcitedefaultmidpunct}
{\mcitedefaultendpunct}{\mcitedefaultseppunct}\relax
\EndOfBibitem
\bibitem[M\"unzer \latin{et~al.}(2010)M\"unzer, Neubauer, Adams, M\"uhlbauer,
  Franz, Jonietz, Georgii, B\"oni, Pedersen, Schmidt, Rosch, and
  Pfleiderer]{Munzer_PRB2010}
M\"unzer,~W.; Neubauer,~A.; Adams,~T.; M\"uhlbauer,~S.; Franz,~C.; Jonietz,~F.;
  Georgii,~R.; B\"oni,~P.; Pedersen,~B.; Schmidt,~M.; Rosch,~A.; Pfleiderer,~C.
  Skyrmion lattice in the doped semiconductor
  ${\text{Fe}}_{1\ensuremath{-}x}{\text{Co}}_{x}\text{Si}$. \emph{Phys. Rev. B}
  \textbf{2010}, \emph{81}, 041203\relax
\mciteBstWouldAddEndPuncttrue
\mciteSetBstMidEndSepPunct{\mcitedefaultmidpunct}
{\mcitedefaultendpunct}{\mcitedefaultseppunct}\relax
\EndOfBibitem
\bibitem[Yu \latin{et~al.}(2011)Yu, Kanazawa, Onose, Kimoto, Zhang, Ishiwata,
  Matsui, and Tokura]{Yu_NatMat2011}
Yu,~X.~Z.; Kanazawa,~N.; Onose,~Y.; Kimoto,~K.; Zhang,~W.~Z.; Ishiwata,~S.;
  Matsui,~Y.; Tokura,~Y. Near room-temperature formation of a skyrmion crystal
  in thin-films of the helimagnet FeGe. \emph{Nature Materials} \textbf{2011},
  \emph{10}, 106--109\relax
\mciteBstWouldAddEndPuncttrue
\mciteSetBstMidEndSepPunct{\mcitedefaultmidpunct}
{\mcitedefaultendpunct}{\mcitedefaultseppunct}\relax
\EndOfBibitem
\bibitem[Yang \latin{et~al.}(2015)Yang, Thiaville, Rohart, Fert, and
  Chshiev]{yang}
Yang,~H.; Thiaville,~A.; Rohart,~S.; Fert,~A.; Chshiev,~M. Anatomy of
  dzyaloshinskii-moriya interaction at Co/Pt interfaces. \emph{Physical review
  letters} \textbf{2015}, \emph{115}, 267210\relax
\mciteBstWouldAddEndPuncttrue
\mciteSetBstMidEndSepPunct{\mcitedefaultmidpunct}
{\mcitedefaultendpunct}{\mcitedefaultseppunct}\relax
\EndOfBibitem
\bibitem[Dup{\'e} \latin{et~al.}(2016)Dup{\'e}, Bihlmayer, B{\"o}ttcher,
  Bl{\"u}gel, and Heinze]{Dupe_NatComm2016}
Dup{\'e},~B.; Bihlmayer,~G.; B{\"o}ttcher,~M.; Bl{\"u}gel,~S.; Heinze,~S.
  Engineering skyrmions in transition-metal multilayers for spintronics.
  \emph{Nature Communications} \textbf{2016}, \emph{7}, 11779\relax
\mciteBstWouldAddEndPuncttrue
\mciteSetBstMidEndSepPunct{\mcitedefaultmidpunct}
{\mcitedefaultendpunct}{\mcitedefaultseppunct}\relax
\EndOfBibitem
\bibitem[Banerjee \latin{et~al.}(2014)Banerjee, Rowland, Erten, and
  Randeria]{Banerjee_PRX2014}
Banerjee,~S.; Rowland,~J.; Erten,~O.; Randeria,~M. Enhanced Stability of
  Skyrmions in Two-Dimensional Chiral Magnets with Rashba Spin-Orbit Coupling.
  \emph{Phys. Rev. X} \textbf{2014}, \emph{4}, 031045\relax
\mciteBstWouldAddEndPuncttrue
\mciteSetBstMidEndSepPunct{\mcitedefaultmidpunct}
{\mcitedefaultendpunct}{\mcitedefaultseppunct}\relax
\EndOfBibitem
\bibitem[Wang \latin{et~al.}(2016)Wang, Zhang, Xu, Peng, Ding, Wang, Hou,
  Zhang, Li, Liu, Wang, Cai, Wang, Li, Hu, Wu, Shen, and
  Zhang]{Wang_AdvMat2016}
Wang,~W. \latin{et~al.}  A Centrosymmetric Hexagonal Magnet with Superstable
  Biskyrmion Magnetic Nanodomains in a Wide Temperature Range of 100–340 K.
  \emph{Advanced Materials} \textbf{2016}, \emph{28}, 6887--6893\relax
\mciteBstWouldAddEndPuncttrue
\mciteSetBstMidEndSepPunct{\mcitedefaultmidpunct}
{\mcitedefaultendpunct}{\mcitedefaultseppunct}\relax
\EndOfBibitem
\bibitem[Lin and Batista(2018)Lin, and Batista]{Lin_PRLL2018}
Lin,~S.-Z.; Batista,~C.~D. Face Centered Cubic and Hexagonal Close Packed
  Skyrmion Crystals in Centrosymmetric Magnets. \emph{Phys. Rev. Lett.}
  \textbf{2018}, \emph{120}, 077202\relax
\mciteBstWouldAddEndPuncttrue
\mciteSetBstMidEndSepPunct{\mcitedefaultmidpunct}
{\mcitedefaultendpunct}{\mcitedefaultseppunct}\relax
\EndOfBibitem
\bibitem[Yu \latin{et~al.}(2012)Yu, Mostovoy, Tokunaga, Zhang, Kimoto, Matsui,
  Kaneko, Nagaosa, and Tokura]{Yu_PNAS2012}
Yu,~X.; Mostovoy,~M.; Tokunaga,~Y.; Zhang,~W.; Kimoto,~K.; Matsui,~Y.;
  Kaneko,~Y.; Nagaosa,~N.; Tokura,~Y. Magnetic stripes and skyrmions with
  helicity reversals. \emph{Proceedings of the National Academy of Sciences}
  \textbf{2012}, \emph{109}, 8856--8860\relax
\mciteBstWouldAddEndPuncttrue
\mciteSetBstMidEndSepPunct{\mcitedefaultmidpunct}
{\mcitedefaultendpunct}{\mcitedefaultseppunct}\relax
\EndOfBibitem
\bibitem[Khanh \latin{et~al.}(2020)Khanh, Nakajima, Yu, Gao, Shibata,
  Hirschberger, Yamasaki, Sagayama, Nakao, Peng, Nakajima, Takagi, Arima,
  Tokura, and Seki]{Khanh_NatNano2020}
Khanh,~N.~D.; Nakajima,~T.; Yu,~X.; Gao,~S.; Shibata,~K.; Hirschberger,~M.;
  Yamasaki,~Y.; Sagayama,~H.; Nakao,~H.; Peng,~L.; Nakajima,~K.; Takagi,~R.;
  Arima,~T.-h.; Tokura,~Y.; Seki,~S. Nanometric square skyrmion lattice in a
  centrosymmetric tetragonal magnet. \emph{Nature Nanotechnology}
  \textbf{2020}, \emph{15}, 444--449\relax
\mciteBstWouldAddEndPuncttrue
\mciteSetBstMidEndSepPunct{\mcitedefaultmidpunct}
{\mcitedefaultendpunct}{\mcitedefaultseppunct}\relax
\EndOfBibitem
\bibitem[Amoroso \latin{et~al.}(2020)Amoroso, Barone, and
  Picozzi]{amoroso2020spontaneous}
Amoroso,~D.; Barone,~P.; Picozzi,~S. Spontaneous skyrmionic lattice from
  anisotropic symmetric exchange in a Ni-halide monolayer. \emph{Nature
  communications} \textbf{2020}, \emph{11}, 1--9\relax
\mciteBstWouldAddEndPuncttrue
\mciteSetBstMidEndSepPunct{\mcitedefaultmidpunct}
{\mcitedefaultendpunct}{\mcitedefaultseppunct}\relax
\EndOfBibitem
\bibitem[Song \latin{et~al.}(2022)Song, Occhialini, Erge{\c{c}}en, Ilyas,
  Amoroso, Barone, Kapeghian, Watanabe, Taniguchi, Botana, Picozzi, Gedik, and
  Comin]{Song_Nature2022}
Song,~Q.; Occhialini,~C.~A.; Erge{\c{c}}en,~E.; Ilyas,~B.; Amoroso,~D.;
  Barone,~P.; Kapeghian,~J.; Watanabe,~K.; Taniguchi,~T.; Botana,~A.~S.;
  Picozzi,~S.; Gedik,~N.; Comin,~R. Evidence for a single-layer van der Waals
  multiferroic. \emph{Nature} \textbf{2022}, \emph{602}, 601--605\relax
\mciteBstWouldAddEndPuncttrue
\mciteSetBstMidEndSepPunct{\mcitedefaultmidpunct}
{\mcitedefaultendpunct}{\mcitedefaultseppunct}\relax
\EndOfBibitem
\bibitem[Qin \latin{et~al.}(2022)Qin, Sayyad, Montblanch, Feuer, Dey, Blei,
  Sailus, Kara, Shen, Yang, Botana, Atature, and Tongay]{Qin_AdvMat2022}
Qin,~Y.; Sayyad,~M.; Montblanch,~A. R.-P.; Feuer,~M. S.~G.; Dey,~D.; Blei,~M.;
  Sailus,~R.; Kara,~D.~M.; Shen,~Y.; Yang,~S.; Botana,~A.~S.; Atature,~M.;
  Tongay,~S. Reaching the Excitonic Limit in 2D Janus Monolayers by In Situ
  Deterministic Growth. \emph{Advanced Materials} \textbf{2022}, \emph{34},
  2106222\relax
\mciteBstWouldAddEndPuncttrue
\mciteSetBstMidEndSepPunct{\mcitedefaultmidpunct}
{\mcitedefaultendpunct}{\mcitedefaultseppunct}\relax
\EndOfBibitem
\bibitem[Liang \latin{et~al.}(2020)Liang, Wang, Du, Hallal, Garcia, Chshiev,
  Fert, and Yang]{Liang_PRB2020}
Liang,~J.; Wang,~W.; Du,~H.; Hallal,~A.; Garcia,~K.; Chshiev,~M.; Fert,~A.;
  Yang,~H. Very large Dzyaloshinskii-Moriya interaction in two-dimensional
  Janus manganese dichalcogenides and its application to realize skyrmion
  states. \emph{Phys. Rev. B} \textbf{2020}, \emph{101}, 184401\relax
\mciteBstWouldAddEndPuncttrue
\mciteSetBstMidEndSepPunct{\mcitedefaultmidpunct}
{\mcitedefaultendpunct}{\mcitedefaultseppunct}\relax
\EndOfBibitem
\bibitem[Xu \latin{et~al.}(2020)Xu, Feng, Prokhorenko, Nahas, Xiang, and
  Bellaiche]{Xu_PRB2020}
Xu,~C.; Feng,~J.; Prokhorenko,~S.; Nahas,~Y.; Xiang,~H.; Bellaiche,~L.
  Topological spin texture in Janus monolayers of the chromium trihalides Cr(I,
  ${X)}_{3}$. \emph{Phys. Rev. B} \textbf{2020}, \emph{101}, 060404\relax
\mciteBstWouldAddEndPuncttrue
\mciteSetBstMidEndSepPunct{\mcitedefaultmidpunct}
{\mcitedefaultendpunct}{\mcitedefaultseppunct}\relax
\EndOfBibitem
\bibitem[Yuan \latin{et~al.}(2020)Yuan, Yang, Cai, Wu, Chen, Yan, and
  Shen]{Yuan_PRB2020}
Yuan,~J.; Yang,~Y.; Cai,~Y.; Wu,~Y.; Chen,~Y.; Yan,~X.; Shen,~L. Intrinsic
  skyrmions in monolayer Janus magnets. \emph{Phys. Rev. B} \textbf{2020},
  \emph{101}, 094420\relax
\mciteBstWouldAddEndPuncttrue
\mciteSetBstMidEndSepPunct{\mcitedefaultmidpunct}
{\mcitedefaultendpunct}{\mcitedefaultseppunct}\relax
\EndOfBibitem
\bibitem[Zhang \latin{et~al.}(2020)Zhang, Xu, Chen, Nahas, Prokhorenko, and
  Bellaiche]{zhang2020emergence}
Zhang,~Y.; Xu,~C.; Chen,~P.; Nahas,~Y.; Prokhorenko,~S.; Bellaiche,~L.
  Emergence of skyrmionium in a two-dimensional CrGe (Se, Te) 3 Janus
  monolayer. \emph{Physical Review B} \textbf{2020}, \emph{102}, 241107\relax
\mciteBstWouldAddEndPuncttrue
\mciteSetBstMidEndSepPunct{\mcitedefaultmidpunct}
{\mcitedefaultendpunct}{\mcitedefaultseppunct}\relax
\EndOfBibitem
\bibitem[{Gilbert}(2004)]{1353448}
{Gilbert},~T.~L. A phenomenological theory of damping in ferromagnetic
  materials. \emph{IEEE Transactions on Magnetics} \textbf{2004}, \emph{40},
  3443--3449\relax
\mciteBstWouldAddEndPuncttrue
\mciteSetBstMidEndSepPunct{\mcitedefaultmidpunct}
{\mcitedefaultendpunct}{\mcitedefaultseppunct}\relax
\EndOfBibitem
\bibitem[Zhang \latin{et~al.}(2021)Zhang, Hou, Wang, and Wu]{Zhang_PRB2021}
Zhang,~B.~H.; Hou,~Y.~S.; Wang,~Z.; Wu,~R.~Q. Tuning Dzyaloshinskii-Moriya
  interactions in magnetic bilayers with a ferroelectric substrate. \emph{Phys.
  Rev. B} \textbf{2021}, \emph{103}, 054417\relax
\mciteBstWouldAddEndPuncttrue
\mciteSetBstMidEndSepPunct{\mcitedefaultmidpunct}
{\mcitedefaultendpunct}{\mcitedefaultseppunct}\relax
\EndOfBibitem
\bibitem[Amoroso \latin{et~al.}(2021)Amoroso, Barone, and
  Picozzi]{Amoroso_Nanomatt2021}
Amoroso,~D.; Barone,~P.; Picozzi,~S. Interplay between Single-Ion and Two-Ion
  Anisotropies in Frustrated 2D Semiconductors and Tuning of Magnetic
  Structures Topology. \emph{Nanomaterials} \textbf{2021}, \emph{11}\relax
\mciteBstWouldAddEndPuncttrue
\mciteSetBstMidEndSepPunct{\mcitedefaultmidpunct}
{\mcitedefaultendpunct}{\mcitedefaultseppunct}\relax
\EndOfBibitem
\bibitem[Yu \latin{et~al.}(2018)Yu, Koshibae, Tokunaga, Shibata, Taguchi,
  Nagaosa, and Tokura]{Yu_Nat2018}
Yu,~X.~Z.; Koshibae,~W.; Tokunaga,~Y.; Shibata,~K.; Taguchi,~Y.; Nagaosa,~N.;
  Tokura,~Y. Transformation between meron and skyrmion topological spin
  textures in a chiral magnet. \emph{Nature} \textbf{2018}, \emph{564},
  95--98\relax
\mciteBstWouldAddEndPuncttrue
\mciteSetBstMidEndSepPunct{\mcitedefaultmidpunct}
{\mcitedefaultendpunct}{\mcitedefaultseppunct}\relax
\EndOfBibitem
\bibitem[Jiang \latin{et~al.}(2018)Jiang, Shan, and Mak]{Jiang_NatMatt2018}
Jiang,~S.; Shan,~J.; Mak,~K.~F. Electric-field switching of two-dimensional van
  der Waals magnets. \emph{Nature Materials} \textbf{2018}, \emph{17},
  406--410\relax
\mciteBstWouldAddEndPuncttrue
\mciteSetBstMidEndSepPunct{\mcitedefaultmidpunct}
{\mcitedefaultendpunct}{\mcitedefaultseppunct}\relax
\EndOfBibitem
\bibitem[Xu \latin{et~al.}(2022)Xu, Ray, Shao, Jiang, Lee, Weber, Goldberger,
  Watanabe, Taniguchi, Muller, Mak, and Shan]{Xu_NatNano2022}
Xu,~Y.; Ray,~A.; Shao,~Y.-T.; Jiang,~S.; Lee,~K.; Weber,~D.; Goldberger,~J.~E.;
  Watanabe,~K.; Taniguchi,~T.; Muller,~D.~A.; Mak,~K.~F.; Shan,~J. Coexisting
  ferromagnetic--antiferromagnetic state in twisted bilayer CrI3. \emph{Nature
  Nanotechnology} \textbf{2022}, \emph{17}, 143--147\relax
\mciteBstWouldAddEndPuncttrue
\mciteSetBstMidEndSepPunct{\mcitedefaultmidpunct}
{\mcitedefaultendpunct}{\mcitedefaultseppunct}\relax
\EndOfBibitem
\bibitem[Akram \latin{et~al.}(2021)Akram, LaBollita, Dey, Kapeghian, Erten, and
  Botana]{Akram_NanoLett2021}
Akram,~M.; LaBollita,~H.; Dey,~D.; Kapeghian,~J.; Erten,~O.; Botana,~A.~S.
  Moir{\'e} Skyrmions and Chiral Magnetic Phases in Twisted CrX3 (X = I, Br,
  and Cl) Bilayers. \emph{Nano Letters} \textbf{2021}, \emph{21},
  6633--6639\relax
\mciteBstWouldAddEndPuncttrue
\mciteSetBstMidEndSepPunct{\mcitedefaultmidpunct}
{\mcitedefaultendpunct}{\mcitedefaultseppunct}\relax
\EndOfBibitem
\bibitem[Akram and Erten(2021)Akram, and Erten]{Akram_PRB2021}
Akram,~M.; Erten,~O. Skyrmions in twisted van der Waals magnets. \emph{Phys.
  Rev. B} \textbf{2021}, \emph{103}, L140406\relax
\mciteBstWouldAddEndPuncttrue
\mciteSetBstMidEndSepPunct{\mcitedefaultmidpunct}
{\mcitedefaultendpunct}{\mcitedefaultseppunct}\relax
\EndOfBibitem
\bibitem[Zhang \latin{et~al.}(2022)Zhang, Wang, Dahlbom, Barros, and
  Batista]{Zhang_arXiv2022}
Zhang,~H.; Wang,~Z.; Dahlbom,~D.; Barros,~K.; Batista,~C.~D. Magnetic skyrmions
  are nanoscale topological textures that have been recently observed in
  different families of quantum magnets. These textures are known as CP.
  \emph{arxiv:2203.15248} \textbf{2022}, \relax
\mciteBstWouldAddEndPunctfalse
\mciteSetBstMidEndSepPunct{\mcitedefaultmidpunct}
{}{\mcitedefaultseppunct}\relax
\EndOfBibitem
\bibitem[Kresse and Furthm\"uller(1996)Kresse, and Furthm\"uller]{vasp}
Kresse,~G.; Furthm\"uller,~J. Efficient iterative schemes for ab initio
  total-energy calculations using a plane-wave basis set. \emph{Phys. Rev. B}
  \textbf{1996}, \emph{54}, 11169--11186\relax
\mciteBstWouldAddEndPuncttrue
\mciteSetBstMidEndSepPunct{\mcitedefaultmidpunct}
{\mcitedefaultendpunct}{\mcitedefaultseppunct}\relax
\EndOfBibitem
\bibitem[Perdew \latin{et~al.}(1996)Perdew, Burke, and Ernzerhof]{pbe}
Perdew,~J.~P.; Burke,~K.; Ernzerhof,~M. Generalized Gradient Approximation Made
  Simple. \emph{Phys. Rev. Lett.} \textbf{1996}, \emph{77}, 3865--3868\relax
\mciteBstWouldAddEndPuncttrue
\mciteSetBstMidEndSepPunct{\mcitedefaultmidpunct}
{\mcitedefaultendpunct}{\mcitedefaultseppunct}\relax
\EndOfBibitem
\bibitem[Bl\"ochl(1994)]{bloch}
Bl\"ochl,~P.~E. Projector augmented-wave method. \emph{Phys. Rev. B}
  \textbf{1994}, \emph{50}, 17953--17979\relax
\mciteBstWouldAddEndPuncttrue
\mciteSetBstMidEndSepPunct{\mcitedefaultmidpunct}
{\mcitedefaultendpunct}{\mcitedefaultseppunct}\relax
\EndOfBibitem
\bibitem[Kresse and Joubert(1999)Kresse, and Joubert]{paw}
Kresse,~G.; Joubert,~D. From ultrasoft pseudopotentials to the projector
  augmented-wave method. \emph{Phys. Rev. B} \textbf{1999}, \emph{59},
  1758--1775\relax
\mciteBstWouldAddEndPuncttrue
\mciteSetBstMidEndSepPunct{\mcitedefaultmidpunct}
{\mcitedefaultendpunct}{\mcitedefaultseppunct}\relax
\EndOfBibitem
\bibitem[Broyden(1970)]{broyden1}
Broyden,~C.~G. {The Convergence of a Class of Double-rank Minimization
  Algorithms 1. General Considerations}. \emph{IMA Journal of Applied
  Mathematics} \textbf{1970}, \emph{6}, 76--90\relax
\mciteBstWouldAddEndPuncttrue
\mciteSetBstMidEndSepPunct{\mcitedefaultmidpunct}
{\mcitedefaultendpunct}{\mcitedefaultseppunct}\relax
\EndOfBibitem
\bibitem[Broyden(1970)]{broyden2}
Broyden,~C.~G. {The Convergence of a Class of Double-rank Minimization
  Algorithms: 2. The New Algorithm}. \emph{IMA Journal of Applied Mathematics}
  \textbf{1970}, \emph{6}, 222--231\relax
\mciteBstWouldAddEndPuncttrue
\mciteSetBstMidEndSepPunct{\mcitedefaultmidpunct}
{\mcitedefaultendpunct}{\mcitedefaultseppunct}\relax
\EndOfBibitem
\bibitem[Baroni \latin{et~al.}(2001)Baroni, de~Gironcoli, Dal~Corso, and
  Giannozzi]{dfpt}
Baroni,~S.; de~Gironcoli,~S.; Dal~Corso,~A.; Giannozzi,~P. Phonons and related
  crystal properties from density-functional perturbation theory. \emph{Rev.
  Mod. Phys.} \textbf{2001}, \emph{73}, 515--562\relax
\mciteBstWouldAddEndPuncttrue
\mciteSetBstMidEndSepPunct{\mcitedefaultmidpunct}
{\mcitedefaultendpunct}{\mcitedefaultseppunct}\relax
\EndOfBibitem
\bibitem[Togo and Tanaka(2015)Togo, and Tanaka]{phonopy}
Togo,~A.; Tanaka,~I. First principles phonon calculations in materials science.
  \emph{Scripta Materialia} \textbf{2015}, \emph{108}, 1--5\relax
\mciteBstWouldAddEndPuncttrue
\mciteSetBstMidEndSepPunct{\mcitedefaultmidpunct}
{\mcitedefaultendpunct}{\mcitedefaultseppunct}\relax
\EndOfBibitem
\bibitem[Dudarev \latin{et~al.}(1998)Dudarev, Botton, Savrasov, Humphreys, and
  Sutton]{dudarev}
Dudarev,~S.~L.; Botton,~G.~A.; Savrasov,~S.~Y.; Humphreys,~C.~J.; Sutton,~A.~P.
  Electron-energy-loss spectra and the structural stability of nickel oxide: An
  LSDA+U study. \emph{Phys. Rev. B} \textbf{1998}, \emph{57}, 1505--1509\relax
\mciteBstWouldAddEndPuncttrue
\mciteSetBstMidEndSepPunct{\mcitedefaultmidpunct}
{\mcitedefaultendpunct}{\mcitedefaultseppunct}\relax
\EndOfBibitem
\bibitem[Xiang \latin{et~al.}(2013)Xiang, Lee, Koo, Gong, and Whangbo]{xiang}
Xiang,~H.; Lee,~C.; Koo,~H.-J.; Gong,~X.; Whangbo,~M.-H. Magnetic properties
  and energy-mapping analysis. \emph{Dalton Transactions} \textbf{2013},
  \emph{42}, 823--853\relax
\mciteBstWouldAddEndPuncttrue
\mciteSetBstMidEndSepPunct{\mcitedefaultmidpunct}
{\mcitedefaultendpunct}{\mcitedefaultseppunct}\relax
\EndOfBibitem
\bibitem[{\v{S}}abani \latin{et~al.}(2020){\v{S}}abani, Bacaksiz, and
  Milo{\v{s}}evi{\'c}]{vsabani}
{\v{S}}abani,~D.; Bacaksiz,~C.; Milo{\v{s}}evi{\'c},~M. Ab initio methodology
  for magnetic exchange parameters: Generic four-state energy mapping onto a
  Heisenberg spin Hamiltonian. \emph{Physical Review B} \textbf{2020},
  \emph{102}, 014457\relax
\mciteBstWouldAddEndPuncttrue
\mciteSetBstMidEndSepPunct{\mcitedefaultmidpunct}
{\mcitedefaultendpunct}{\mcitedefaultseppunct}\relax
\EndOfBibitem
\bibitem[Mentink \latin{et~al.}(2010)Mentink, Tretyakov, Fasolino, Katsnelson,
  and Rasing]{Mentink_2010}
Mentink,~J.~H.; Tretyakov,~M.~V.; Fasolino,~A.; Katsnelson,~M.~I.; Rasing,~T.
  Stable and fast semi-implicit integration of the stochastic
  Landau{\textendash}Lifshitz equation. \emph{Journal of Physics: Condensed
  Matter} \textbf{2010}, \emph{22}, 176001\relax
\mciteBstWouldAddEndPuncttrue
\mciteSetBstMidEndSepPunct{\mcitedefaultmidpunct}
{\mcitedefaultendpunct}{\mcitedefaultseppunct}\relax
\EndOfBibitem
\bibitem[Luttinger and Tisza(1946)Luttinger, and Tisza]{Luttinger_PR1946}
Luttinger,~J.~M.; Tisza,~L. Theory of Dipole Interaction in Crystals.
  \emph{Phys. Rev.} \textbf{1946}, \emph{70}, 954--964\relax
\mciteBstWouldAddEndPuncttrue
\mciteSetBstMidEndSepPunct{\mcitedefaultmidpunct}
{\mcitedefaultendpunct}{\mcitedefaultseppunct}\relax
\EndOfBibitem
\bibitem[Hayami \latin{et~al.}(2016)Hayami, Lin, and
  Batista]{PhysRevB.93.184413}
Hayami,~S.; Lin,~S.-Z.; Batista,~C.~D. Bubble and skyrmion crystals in
  frustrated magnets with easy-axis anisotropy. \emph{Phys. Rev. B}
  \textbf{2016}, \emph{93}, 184413\relax
\mciteBstWouldAddEndPuncttrue
\mciteSetBstMidEndSepPunct{\mcitedefaultmidpunct}
{\mcitedefaultendpunct}{\mcitedefaultseppunct}\relax
\EndOfBibitem
\bibitem[Batista \latin{et~al.}(2016)Batista, Lin, Hayami, and
  Kamiya]{Batista_2016}
Batista,~C.~D.; Lin,~S.-Z.; Hayami,~S.; Kamiya,~Y. Frustration and chiral
  orderings in correlated electron systems. \emph{Reports on Progress in
  Physics} \textbf{2016}, \emph{79}, 084504\relax
\mciteBstWouldAddEndPuncttrue
\mciteSetBstMidEndSepPunct{\mcitedefaultmidpunct}
{\mcitedefaultendpunct}{\mcitedefaultseppunct}\relax
\EndOfBibitem
\bibitem[Nagaosa and Tokura(2013)Nagaosa, and Tokura]{nagaosa2013topological}
Nagaosa,~N.; Tokura,~Y. Topological properties and dynamics of magnetic
  skyrmions. \emph{Nature nanotechnology} \textbf{2013}, \emph{8},
  899--911\relax
\mciteBstWouldAddEndPuncttrue
\mciteSetBstMidEndSepPunct{\mcitedefaultmidpunct}
{\mcitedefaultendpunct}{\mcitedefaultseppunct}\relax
\EndOfBibitem
\bibitem[Berg and L\"uscher(1981)Berg, and L\"uscher]{BERG1981412}
Berg,~B.; L\"uscher,~M. Definition and statistical distributions of a
  topological number in the lattice O(3) $\sigma$-model. \emph{Nuclear Physics
  B} \textbf{1981}, \emph{190}, 412--424\relax
\mciteBstWouldAddEndPuncttrue
\mciteSetBstMidEndSepPunct{\mcitedefaultmidpunct}
{\mcitedefaultendpunct}{\mcitedefaultseppunct}\relax
\EndOfBibitem
\end{mcitethebibliography}


\providecommand{\latin}[1]{#1}
\makeatletter
\providecommand{\doi}
  {\begingroup\let\do\@makeother\dospecials
  \catcode`\{=1 \catcode`\}=2 \doi@aux}
\providecommand{\doi@aux}[1]{\endgroup\texttt{#1}}
\makeatother
\providecommand*\mcitethebibliography{\thebibliography}
\csname @ifundefined\endcsname{endmcitethebibliography}
  {\let\endmcitethebibliography\endthebibliography}{}
\begin{mcitethebibliography}{4}
\providecommand*\natexlab[1]{#1}
\providecommand*\mciteSetBstSublistMode[1]{}
\providecommand*\mciteSetBstMaxWidthForm[2]{}
\providecommand*\mciteBstWouldAddEndPuncttrue
  {\def\EndOfBibitem{\unskip.}}
\providecommand*\mciteBstWouldAddEndPunctfalse
  {\let\EndOfBibitem\relax}
\providecommand*\mciteSetBstMidEndSepPunct[3]{}
\providecommand*\mciteSetBstSublistLabelBeginEnd[3]{}
\providecommand*\EndOfBibitem{}
\mciteSetBstSublistMode{f}
\mciteSetBstMaxWidthForm{subitem}{(\alph{mcitesubitemcount})}
\mciteSetBstSublistLabelBeginEnd
  {\mcitemaxwidthsubitemform\space}
  {\relax}
  {\relax}

\bibitem[Xiang \latin{et~al.}(2013)Xiang, Lee, Koo, Gong, and Whangbo]{xiang}
Xiang,~H.; Lee,~C.; Koo,~H.-J.; Gong,~X.; Whangbo,~M.-H. Magnetic properties
  and energy-mapping analysis. \emph{Dalton Transactions} \textbf{2013},
  \emph{42}, 823--853\relax
\mciteBstWouldAddEndPuncttrue
\mciteSetBstMidEndSepPunct{\mcitedefaultmidpunct}
{\mcitedefaultendpunct}{\mcitedefaultseppunct}\relax
\EndOfBibitem
\bibitem[{\v{S}}abani \latin{et~al.}(2020){\v{S}}abani, Bacaksiz, and
  Milo{\v{s}}evi{\'c}]{vsabani}
{\v{S}}abani,~D.; Bacaksiz,~C.; Milo{\v{s}}evi{\'c},~M. Ab initio methodology
  for magnetic exchange parameters: Generic four-state energy mapping onto a
  Heisenberg spin Hamiltonian. \emph{Physical Review B} \textbf{2020},
  \emph{102}, 014457\relax
\mciteBstWouldAddEndPuncttrue
\mciteSetBstMidEndSepPunct{\mcitedefaultmidpunct}
{\mcitedefaultendpunct}{\mcitedefaultseppunct}\relax
\EndOfBibitem
\bibitem[Amoroso \latin{et~al.}(2020)Amoroso, Barone, and
  Picozzi]{amoroso2020spontaneous}
Amoroso,~D.; Barone,~P.; Picozzi,~S. Spontaneous skyrmionic lattice from
  anisotropic symmetric exchange in a Ni-halide monolayer. \emph{Nature
  communications} \textbf{2020}, \emph{11}, 1--9\relax
\mciteBstWouldAddEndPuncttrue
\mciteSetBstMidEndSepPunct{\mcitedefaultmidpunct}
{\mcitedefaultendpunct}{\mcitedefaultseppunct}\relax
\EndOfBibitem
\end{mcitethebibliography}
	
\end{document}




\subsection{Molecular dynamic simulations}
\begin{figure}
\renewcommand{\figurename}{Figure}
\renewcommand{\thefigure}{S\arabic{figure}}
\centering
\includegraphics[width=16.0cm]{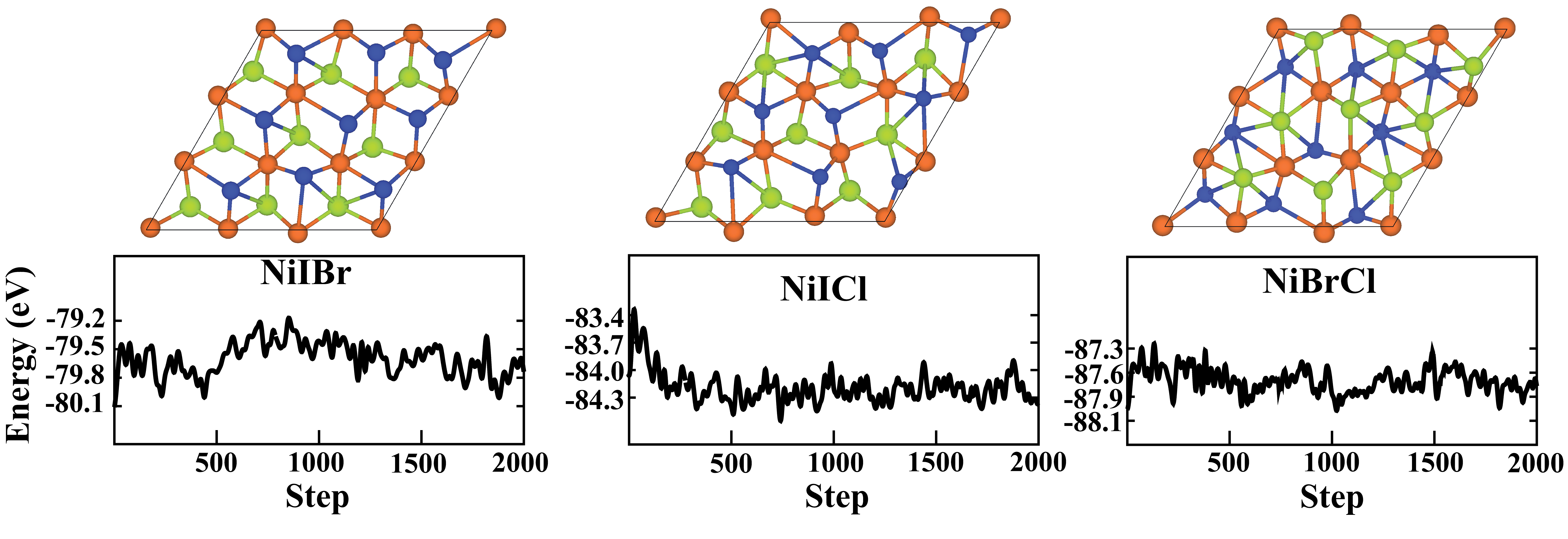}
\caption{Top panels: Snapshots of AIMD simulations for all NiXY monolayers after 4ps. Bottom: Variation of the total energy  against simulation time step (2000x2fs=4ps) for the AIMD simulations performed at 300 K.}
\label{fig:figS1}
\end{figure}

Ab-initio molecular dynamic simulations (AIMD) were performed within a 3$\times$3$\times$1 supercell and a 5$\times$5$\times$1 k-point grid using a micro-canonical ensemble method. Newton’s equation of motion was integrated through the Verlet algorithm with time steps of 2 fs. 
All structures considered in this study are kept at 300 K for 4 ps. The top panels of Figure \ref{fig:figS1} are the snapshot of AIMD simulations. In the course of MD iterations lasting 4 ps, the structure of NiXY monolayers (belonging to the space group P${3}$m1) remains unaltered. The bottom panels of Fig. \ref{fig:figS1} show the variation of total energy with respect to the number of time steps at 300 K for all NiXY monolayers. 

In addition to AIMD simulations and the phonon calculations presented in the main text, to determine the relative energetic stabilities of the fully relaxed NiXY, the average cohesive (E$_c$) energies were also calculated. The average cohesive energy (per atom) for NiXY monolayers was obtained by using the expression, E$_c$= (E$_{Ni}$ + E$_X$ + E$_Y$ - E$_{NiXY}$)/3. Here, E$_{NiXY}$ is the total energy of the optimized  NiXY monolayer, E$_{Ni}$, E$_X$, and E$_Y$ represent the single-atom energy of Ni, and halide (X, Y) atoms, respectively. The average cohesive energy calculations were obtained for a ferromagnetic state within PBE. The calculated E$_c$ of NiI$_2$, NiIBr, NiICl, and NiBrCl is 2.72, 2.86, 2.99, and 3.20 eV, respectively, which indicates that they all correspond to energetically stable structures. Within PBE + $U$ ( $U$=1eV), the energies are 5$\%$ smaller and still positive, indicating the stability of the monolayers within PBE+ $U$ as well.

\subsection{Electronic band structures}

\begin{figure}[H]
\renewcommand{\figurename}{Figure}
\renewcommand{\thefigure}{S\arabic{figure}}
\centering
\includegraphics[width=16.0cm]{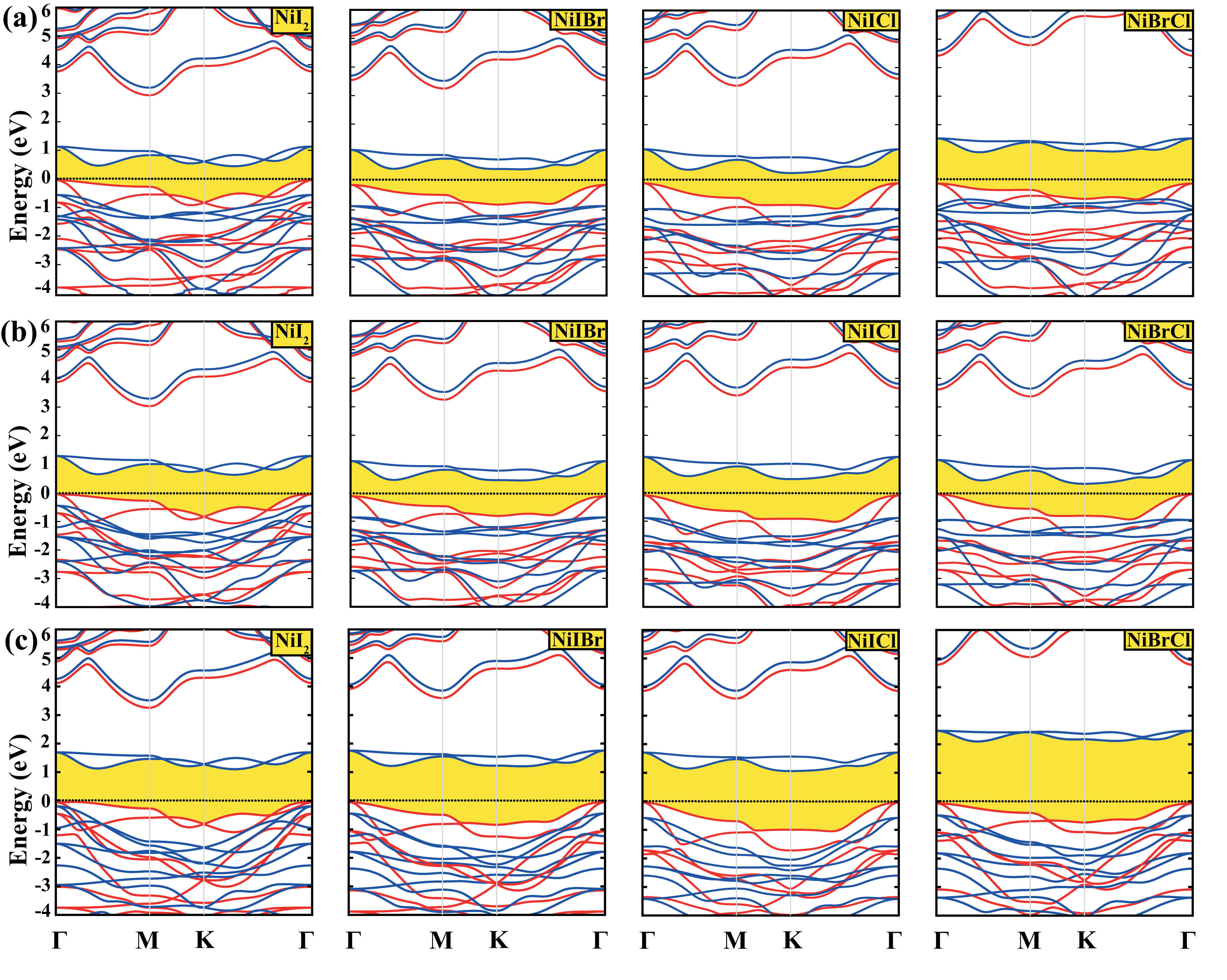}
\caption{Evolution of the band structure for NiI$_2$ and NiXY monolayers at different  $U$ values ( $U$= 1 (top panels), 2 (middle panels) and 5 eV (bottom panels)) shown along high-symmetry directions. Blue and red colors represent spin-up and spin-down states,  respectively.}
\label{fig:figS3}
\end{figure}

We now discuss the electronic band structures of all NiXY and NiI$_2$ monolayers within GGA+ $U$ calculated in a ferromagnetic state using  $U$= 1, 2, and 5 eV (see  Table \ref{tableS2} with the calculated band gap values and Fig. \ref{fig:figS3} with the corresponding band structures at different  $U$s).  For NiI$_2$, an insulating state is obtained with a band gap value that changes from 0.5 eV ( $U$= 1 eV) to 1.15 eV ( $U$= 5 eV) 
. The states near the valence band maximum (VBM) and the conduction band minimum (CBM) are mainly composed of Ni-3$d$, and I-5$p$ hybridized orbitals.
 
For all NiXY monolayers a gap also opens up for the range of  $U$ values studied here. In comparison  to the NiI$_2$ monolayer, the band gap of NiIBr and NiBrCl monolayers increases (from 0.5 eV for NiI$_2$, to 0.55 eV for NiIBr to 1.09 eV for NiBrCl at  $U$ = 1 eV), while for NiICl it slightly decreases (from 0.5 eV to 0.35 eV at  $U$= 1 eV). Similarly to the NiI$_2$ monolayer, the Ni-3$d$ and I(Br or Cl)-5$p$ hybrid orbitals provide the dominant contribution to the states near the VBM and CBM for the NiXY monolayers. For larger  $U$ values, the main characteristics of the band structure remain, with the band gap simply increasing with  $U$, as expected.


\begin{table}[H]
\begin{tabular}{ccccc}
\multicolumn{1}{l}{}    & \multicolumn{1}{l}{} & $E_{\uparrow}$ (eV) & $E_{\downarrow}$(eV) & $E_{gap}$(eV) \\ \hline
\multirow{3}{*}{ NiI$_2$}   &  $U$=1 & 1.00 & 2.96 & 0.50 \\
                        &  $U$=2 & 1.07 & 3.08 & 0.68 \\
                        &  $U$=5 & 1.30 & 3.30 & 1.15 \\ \hline
\multirow{3}{*}{NiIBr}  &  $U$=1 & 1.29 & 3.37 & 0.55 \\
                        &  $U$=2 & 1.39 & 3.46 & 0.74 \\
                        &  $U$=5 & 1.65 & 3.64 & 1.26 \\ \hline
\multirow{3}{*}{NiICl}  &  $U$=1 & 1.25 & 3.42 & 0.35 \\
                        &  $U$=2 & 1.35 & 3.49 & 0.56 \\
                        &  $U$=5 & 1.64 & 3.64 & 1.10 \\ \hline
\multirow{3}{*}{NiBrCl} &  $U$=1 & 1.64 & 4.47 & 1.09 \\
                        &  $U$=2 & 2.06 & 4.61 & 1.37 \\
                        &  $U$=5 & 2.59 & 4.85 & 2.17 \\ \hline
\end{tabular}
\renewcommand{\tablename}{Table}
\renewcommand{\thetable}{S\arabic{table}}
\captionof{table}{Spin-up gaps ($E_{\uparrow}$), spin-down gaps ($E_{\downarrow}$) and band gaps ($E_{gap}$) are listed for NiI$_2$ and NiXY monolayers. As the U values increase, the band gap values increase, as expected.}
\label{tableS2}
\end{table}

\subsection{Magnetic constants and four-state method}

As mentioned in the main text, we calculated the magnetic parameters via the four-state
energy mapping method,  explained in detail in Ref.\citenum{xiang} 
performing noncollinear DFT calculations with SOC and
constraining the direction of the magnetic moments. 
By means of this method, all the elements of the exchange tensor for
a chosen magnetic pair can be obtained, thus gaining direct access to the symmetric anisotropic
exchange part and the antisymmetric anisotropic part
(DMI) of the exchange. 

\begin{figure}[H]
\renewcommand{\figurename}{Figure}
\renewcommand{\thefigure}{S\arabic{figure}}
\centering
\includegraphics[width=7.0cm]{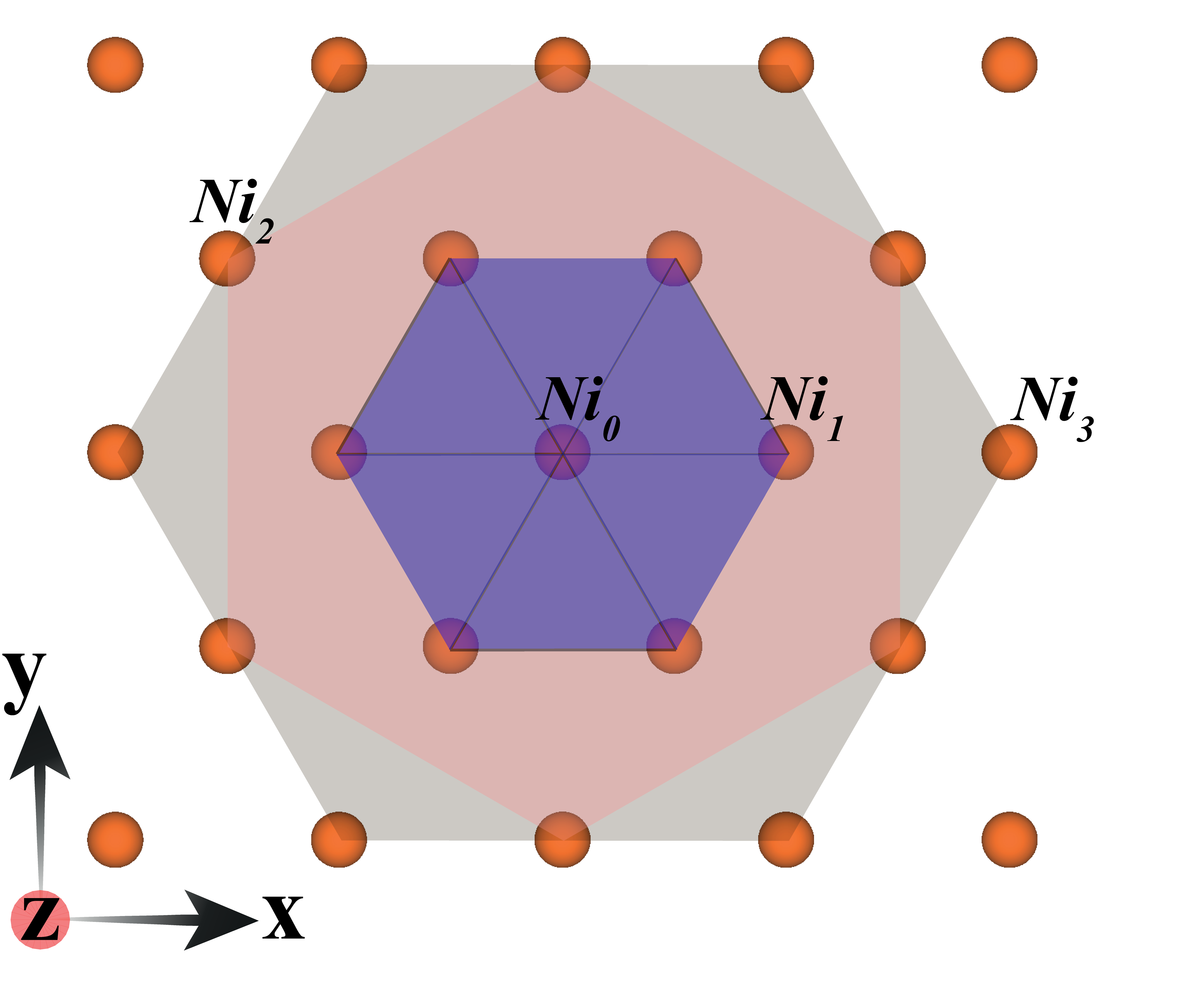}
\caption{Ni atoms are shown for  NiI$_{2}$ and NiXY monolayers (orange spheres). Ni$_{0}$ is the central atom. The Ni atoms at the corners of the purple, pink, and grey hexagons are the first, second, and third nearest neighbors of Ni$_{0}$.  The Ni$_{0}$-Ni$_{1}$ pair (along the x axis) is the one selected for the first nearest-neighbor interaction. The Ni$_{0}$-Ni$_{2}$ pair shows the Ni atom pair selected for the second nearest-neighbor interaction, while the Ni$_{0}$-Ni$_{3}$ pair shows the Ni pair selected for the third nearest-neighbor interaction.}
\label{fig:figS4}
\end{figure}

In particular, we performed calculations on the magnetic pair parallel to the x direction Ni$_0$- Ni$_1$ (see Fig. \ref{fig:figS4}). The interaction between the five other nearest-neighbor pairs can
be evaluated via the three-fold rotational symmetry. In NiI$_2$ the tensor is
symmetric whereas for Janus monolayers there is an anti-symmetric (DMI) contribution.
As mentioned in the methods, we performed calculations of the SIA, first nearest-neighbor (NN) and second-NN interactions using a 5 × 4 × 1 supercell. A 6 × 3 × 1 supercell was used to estimate 
the third-NN interaction instead. 

Given that the four-state energy calculations are noncolinear (with SOC included), the spin configurations for each Ni atom are specified in the x-y-z direction (in cartesian coordinates).  To obtain the J$_{\rm ij}$ terms, the DFT energies calculated from the four different states are used in the equation J$_{\rm ij}$ = (E$_{\rm 1}$ + E$_{\rm 4}$ - E$_{\rm 2}$ - E$_{\rm 3}$) /2S$^{\rm 2}$ (for a detailed derivation, see Ref.\citenum{vsabani}). 
The spin maps used to calculate the NN J$_{\rm ij}$ terms are shown in Table \ref{tableS3} (the same maps are used for the second and third nearest-neighbor interactions). To obtain the SIA, the four state method is applied on a single Ni atom: (0 0 S), (0 0 -S),(S 0 0), (-S 0 0) while other Ni ions and also the halide
atoms have no spin (0 0 0). To obtain the A$_{\rm zz}$-A$_{\rm xx}$ term, the DFT energies calculated from four different states are used in the equation A$_{\rm zz}$-A$_{\rm xx}$ = (E$_{\rm 1}$ + E$_{\rm 2}$ - E$_{\rm 3}$ - E$_{\rm 4}$) /2S$^{\rm 2}$  (for a detailed derivation, once again follow Ref.\cite{vsabani}). 

\begin{table}[H]
\renewcommand{\tablename}{Table}
\renewcommand{\thetable}{S\arabic{table}}
\resizebox{\textwidth}{!}{%
\begin{tabular}{cccccccccc}
                            & Ni$_0$      & Ni$_1$      & Other Ni's                     & Ni$_0$     & Ni$_1$      & Other Ni's                     & Ni$_0$     &Ni$_1$      & Other Ni's                     \\ \cline{2-10} 
\multicolumn{1}{c|}{States} & \multicolumn{3}{c|}{J$_{\rm xx}$}                      & \multicolumn{3}{c|}{J$_{\rm xy}$}                      & \multicolumn{3}{c|}{J$_{\rm xz}$}                      \\
\multicolumn{1}{c|}{1}     & (S,0,0)  & (S,0,0)  & \multicolumn{1}{c|}{(0,0,S)} & (S,0,0)  & (0,S,0)  & \multicolumn{1}{c|}{(0,0,S)} & (S,0,0)  & (0,0,S)  & \multicolumn{1}{c|}{(0,S,0)} \\
\multicolumn{1}{c|}{2}     & (S,0,0)  & (-S,0,0) & \multicolumn{1}{c|}{(0,0,S)} & (S,0,0)  & (0,-S,0) & \multicolumn{1}{c|}{(0,0,S)} & (S,0,0)  & (0,0,-S) & \multicolumn{1}{c|}{(0,S,0)} \\
\multicolumn{1}{c|}{3}     & (-S,0,0) & (S,0,0)  & \multicolumn{1}{c|}{(0,0,S)} & (-S,0,0) & (0,S,0)  & \multicolumn{1}{c|}{(0,0,S)} & (-S,0,0) & (0,0,S)  & \multicolumn{1}{c|}{(0,S,0)} \\
\multicolumn{1}{c|}{4}     & (-S,0,0) & (-S,0,0) & \multicolumn{1}{c|}{(0,0,S)} & (-S,0,0) & (0,-S,0) & \multicolumn{1}{c|}{(0,0,S)} & (-S,0,0) & (0,0,-S) & \multicolumn{1}{c|}{(0,S,0)} \\ \cline{2-10} 
\multicolumn{1}{c|}{}       & \multicolumn{3}{c|}{J$_{\rm yx}$}                      & \multicolumn{3}{c|}{J$_{\rm yy}$}                      & \multicolumn{3}{c|}{J$_{\rm yz}$}                      \\
\multicolumn{1}{c|}{1}     & (0,S,0)  & (S,0,0)  & \multicolumn{1}{c|}{(0,0,S)} & (0,S,0)  & (0,S,0)  & \multicolumn{1}{c|}{(0,0,S)} & (0,S,0)  & (0,0,S)  & \multicolumn{1}{c|}{(S,0,0)} \\
\multicolumn{1}{c|}{2}     & (0,S,0)  & (-S,0,0) & \multicolumn{1}{c|}{(0,0,S)} & (0,S,0)  & (0,-S,0) & \multicolumn{1}{c|}{(0,0,S)} & (0,S,0)  & (0,0,-S) & \multicolumn{1}{c|}{(S,0,0)} \\
\multicolumn{1}{c|}{3}     & (0,-S,0) & (S,0,0)  & \multicolumn{1}{c|}{(0,0,S)} & (0,-S,0) & (0,S,0)  & \multicolumn{1}{c|}{(0,0,S)} & (0,-S,0) & (0,0,S)  & \multicolumn{1}{c|}{(S,0,0)} \\
\multicolumn{1}{c|}{4}     & (0,-S,0) & (-S,0,0) & \multicolumn{1}{c|}{(0,0,S)} & (0,-S,0) & (0,-S,0) & \multicolumn{1}{c|}{(0,0,S)} & (0,-S,0) & (0,0,-S) & \multicolumn{1}{c|}{(S,0,0)} \\ \cline{2-10} 
\multicolumn{1}{c|}{}       & \multicolumn{3}{c|}{J$_{\rm zx}$}                      & \multicolumn{3}{c|}{J$_{\rm zy}$}                      & \multicolumn{3}{c|}{J$_{\rm zz}$}                      \\
\multicolumn{1}{c|}{1}     & (0,0,S)  & (S,0,0)  & \multicolumn{1}{c|}{(0,S,0)} & (0,0,S)  & (0,S,0)  & \multicolumn{1}{c|}{(0,S,0)} & (0,0,S)  & (0,0,S)  & \multicolumn{1}{c|}{(S,0,0)} \\
\multicolumn{1}{c|}{2}     & (0,0,S)  & (-S,0,0) & \multicolumn{1}{c|}{(0,S,0)} & (0,0,S)  & (0,-S,0) & \multicolumn{1}{c|}{(0,S,0)} & (0,0,S)  & (0,0,-S) & \multicolumn{1}{c|}{(S,0,0)} \\
\multicolumn{1}{c|}{3}     & (0,0,-S) & (S,0,0)  & \multicolumn{1}{c|}{(0,S,0)} & (0,0,-S) & (0,S,0)  & \multicolumn{1}{c|}{(0,S,0)} & (0,0,-S) & (0,0,S)  & \multicolumn{1}{c|}{(S,0,0)} \\
\multicolumn{1}{c|}{4}     & (0,0,-S) & (-S,0,0) & \multicolumn{1}{c|}{(0,S,0)} & (0,0,-S) & (0,-S,0) & \multicolumn{1}{c|}{(0,S,0)} & (0,0,-S) & (0,0,-S) & \multicolumn{1}{c|}{(S,0,0)} \\ \cline{2-10} 
\end{tabular}%
}
\captionof{table}{Spin maps used in the four-state method  to obtain each J$_{\rm ij}$ term.}
\label{tableS3}
\end{table}


The exchange coupling tensor considered in the main text for NiI$_2$ for a Ni-Ni pair whose bonding vector is chosen
parallel to the cartesian axis x (as shown in Fig. \ref{fig:figS4}) is: 

\begin{equation}
J^{1 NiI_2 (0^{\circ})}=
\begin{pmatrix}\label{J}
J_{xx} & 0 & 0 \\
0 & J_{yy} & J_{yz} \\
0 & J_{yz} & J_{zz} 
\end{pmatrix} 
\end{equation} 

wherein off-diagonal terms except for J$_{yz}$ are nominally zero.

The corresponding tensor for the symmetry-equivalent pairs rotated 120 $^{\circ}$
can be deduced exploiting the three-fold rotational symmetry, leading to: 


\begin{equation}
J^{NiI_2}(120^{\circ})=
  \begin{pmatrix}
  \frac{1}{4}(J_{xx} + 3J_{yy}) & - \frac{\sqrt{3}}{4}(J_{xx} - J_{yy}) & -\frac{\sqrt{3}}{2}J_{yz} \\ \\
-\frac{\sqrt{3}}{4}(J_{xx} - J_{yy}) & \frac{1}{4}(3J_{xx} + J_{yy}) & -\frac{1}{2}J_{yz} \\ \\
-\frac{\sqrt{3}}{2}J_{yz}  & -\frac{1}{2}J_{yz}  & J_{zz}
  \end{pmatrix}
\end{equation}\label{J120}

For the NiI$_2$ monolayer,  the J$^{NiI_2}$ (Eq. \ref{J}) tensor, can be analytically diagonalized (for detailed information, see Ref. \citenum{amoroso2020spontaneous}). 

For NiXY monolayers, the off-diagonal terms of the J tensor are non-zero (non-symmetrical off-diagonal terms are related to the DMI), and the symmetry-equivalent pairs have instead a tensor that looks as follows:


\begin{equation}
J^{NiXY}(120^{\circ})= \mbox{\tiny$
\begin{pmatrix} 
\frac{1}{4}(J_{xx} +3J_{yy} +\sqrt{3}J_{yx} +\sqrt{3}J_{xy}) & -\frac{1}{4}(\sqrt{3}J_{xx} - \sqrt{3}J_{yy} + 3J_{yx} - J_{xy} ) & -\frac{1}{2}(\sqrt{3}J_{yz}+J_{xz}) \\ \\
-\frac{1}{4}(\sqrt{3}J_{xx} - \sqrt{3}J_{yy} - J_{yx} + 3J_{xy} ) & \frac{1}{4}(3J_{xx} +J_{yy} -\sqrt{3}J_{yx} -\sqrt{3}J_{xy}) & \frac{1}{2}(\sqrt{3}J_{xz}-J_{yz}) \\ \\
-\frac{1}{2}(J_{zx}+\sqrt{3}J_{zy})  & \frac{1}{2}(\sqrt{3}J_{zx}-J_{zy})  & J_{zz} 
\end{pmatrix}$} 
\end{equation}\label{J2120}



As mentioned in the main text, the exchange tensor we discuss therein is expressed in terms of the
cartesian \{x, y, z\} basis, with x is parallel to the
Ni-Ni bonding vector, as explained above.
In order to better visualize the role of the two-site
anisotropy, it is useful to express the interaction within the local
principal-axes basis  \{$\alpha, \beta, \gamma$\} which diagonalize the exchange
tensor, as detailed in Ref. \citenum{amoroso2020spontaneous}. 
The corresponding mumerical eigenvalues for the first NN exchange tensor are listed in Table \ref{tableS5} for  $U$=1 eV for NiI$_2$ and NiXY monolayers.  Note that the J tensor (for NiXY monolayers) is not symmetrical, and eigenvalues and eigenvectors are complex numbers. Table \ref{tableS4}  shows the only real part of eigenvalues in terms of local orthogonal basis \{$\alpha, \beta, \gamma$\}. For NiXY monolayers $\lambda_\alpha$= $\lambda_\beta$ but for NiI$_2$, $\lambda_\alpha$ is slightly different from  $\lambda_\beta$ . 

\begin{table}[H]
\begin{tabular}{llll}
        & $\lambda_\alpha$     & $\lambda_\beta$    & $\lambda_\gamma$     \\ \hline
NiI$_2$ & -8.83 & -8.72 & -5.05 \\ \hline
NiIBr   & -6.82 & -6.82 & -4.70 \\ \hline
NiICl   & -4.71 & -4.71 & -2.70 \\ \hline
NiBrCl  & -5.93 & -5.93 & -5.74 \\ \hline
\end{tabular}
\renewcommand{\tablename}{Table}
\renewcommand{\thetable}{S\arabic{table}}
\captionof{table}{Eigenvalues of the first nearest neighbor exchange tensor within the local coordinate system  \{$\alpha, \beta, \gamma$\} for NiXY and NiI$_2$ monolayers (in meV) for calculations performed at  $U$=1 eV.}
\label{tableS5}
\end{table}

Finally, to answer how different  $U$ values affect the  magnetic constants, we derived them at  $U$=1, 2, and 5 eV (see Table \ref{tableS4}). The J$_{\rm ij}$ terms depend on the  $U$ values, as expected, while important ratios (J$_{\rm yz}$/J$^{\rm 1iso}$ , J$^{\rm 3iso}$/ J$^{\rm 1iso}$)  remain approximately constant with U (see main text for further details).  

\begin{table}[H]
\begin{tabular}{llllllll}
                          &     & J$^{\rm 1iso}$ & J$^{\rm 2iso}$& J$^{\rm 3iso}$& SIA   & J$^{\rm 3iso}$/J$^{\rm 1iso}$ & J$_{\rm yz}$/J$^{\rm 1iso}$ \\ \hline
\multirow{3}{*}{NiI$_2$}  &  $U$=1 & -7.53     &  0.02     & 6.15      & 0.64 & -0.82                        & -0.22                  \\
                          &  $U$=2 & -6.41     & -0.02     & 4.98      & 0.49 & -0.78                        & -0.20                  \\
                          &  $U$=5 & -3.76     & -0.07     & 2.57      & 0.04 & -0.68                        & -0.15                  \\ \hline
\multirow{3}{*}{NiIBr}    &  $U$=1 & -6.12     & -0.01     & 5.48      & 0.12 & -0.90                        & -0.17                  \\
                          &  $U$=2 & -5.16     & 0.02      & 4.41      & 0.10 & -0.85                        & -0.16                  \\
                          &  $U$=5 & -2.95     & -0.03     & 2.24      & 0.05 & -0.76                        & -0.12                  \\ \hline
\multirow{3}{*}{NiICl}    &  $U$=1 & -4.04     & -0.33     & 6.14      & 0.00 & -1.52                        & -0.27                  \\
                          &  $U$=2 & -3.56     & -0.15     & 4.94      & 0.00 & -1.39                        & -0.24                  \\
                          &  $U$=5 & -2.23     & -0.03     & 3.00      & 0.01 & -1.34                        & -0.17                  \\ \hline
\multirow{3}{*}{NiBrCl}   &  $U$=1 & -5.87     & -0.08     & 2.76      & 0.00 & -0.47                        & -0.02                  \\
                          &  $U$=2 & -4.79     & -0.04     & 2.18      & 0.00 & -0.45                        & -0.02                  \\
                          &  $U$=5 & -2.54     & -0.03     & 1.06      & 0.00 & -0.42                        & -0.01                  \\ \hline
\end{tabular}
\renewcommand{\tablename}{Table}
\renewcommand{\thetable}{S\arabic{table}}
\captionof{table}{Isotropic first, second, and third nearest-neighbor interactions, SIA, J$^{\rm 3iso}$/J$^{\rm 1iso}$ and J$_{\rm yz}$/J$^{\rm 1iso}$ with different  $U$ values for NiI$_2$ and NiXY monolayers. All terms are listed in meV. Negative and positive values of J$^{\rm niso}$ are ferromagnetic and antiferromagnetic, respectively. Positive values of SIA indicate easy-plane anisotropy. SIA values of less than 5 $\mu$eV are listed as zero.}
\label{tableS4}
\end{table}

\subsection{Atomistic simulations for NiI$_2$}
\begin{figure}[H]
\renewcommand{\figurename}{Figure}
\renewcommand{\thefigure}{S\arabic{figure}}
\centering
\includegraphics[width=8.4cm]{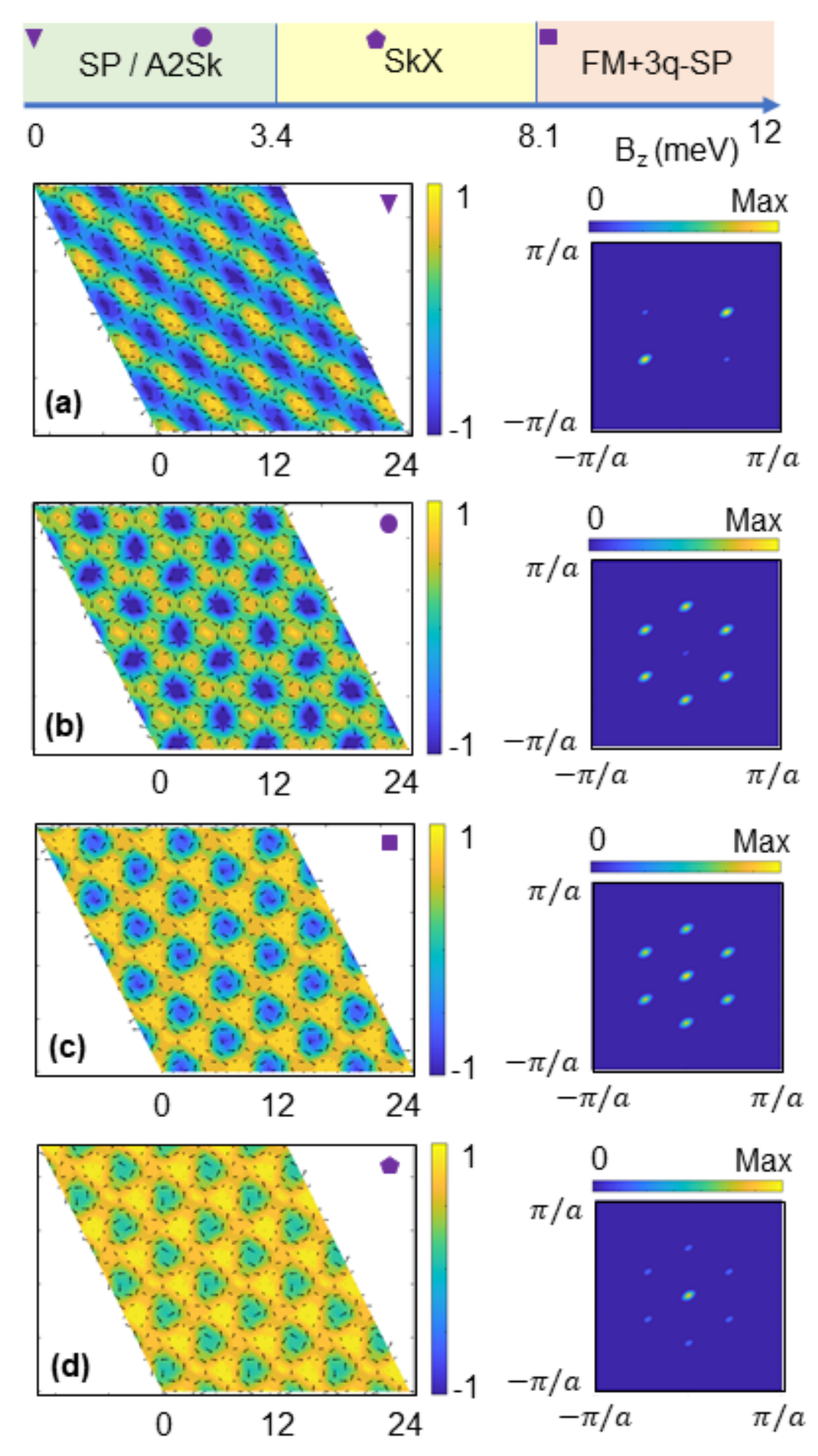}
\caption{Phase diagram of NiI$_2$ as a function external magnetic field B. Magnetization texture and spin structure factor for (a) $B=0 \rm~ meV$, (b) $B=2.8 \rm ~meV$, (c) $B=5.6 \rm ~meV$. (d) $B=8.2 ~\rm meV$.}
\label{fig_NiI2_phasediagram}
\end{figure}

We present the phase diagram of NiI$_2$ as a function of $B$ in Fig. \ref{fig_NiI2_phasediagram}. For low magnetic field, we find A2Sk and Sp are quasidegenerate. At $B =3.4$ meV, we get a SkX phase of Bloch skyrmions which survives up to $B=8.1$ meV. As the magnetic field is further increased, a FM+3q-SP phase is obtained with six peaks in the spin structure factor. This phase adiabatically connects with a fully-polarized FM at larger fields.

\subsection{Topological charge densities}
In this section, we present the topological charge densities for the magnetic textures that are shown in the main text.
\begin{figure}[H]
\renewcommand{\figurename}{Figure}
\renewcommand{\thefigure}{S\arabic{figure}}
\centering
\includegraphics[width=8.4cm]{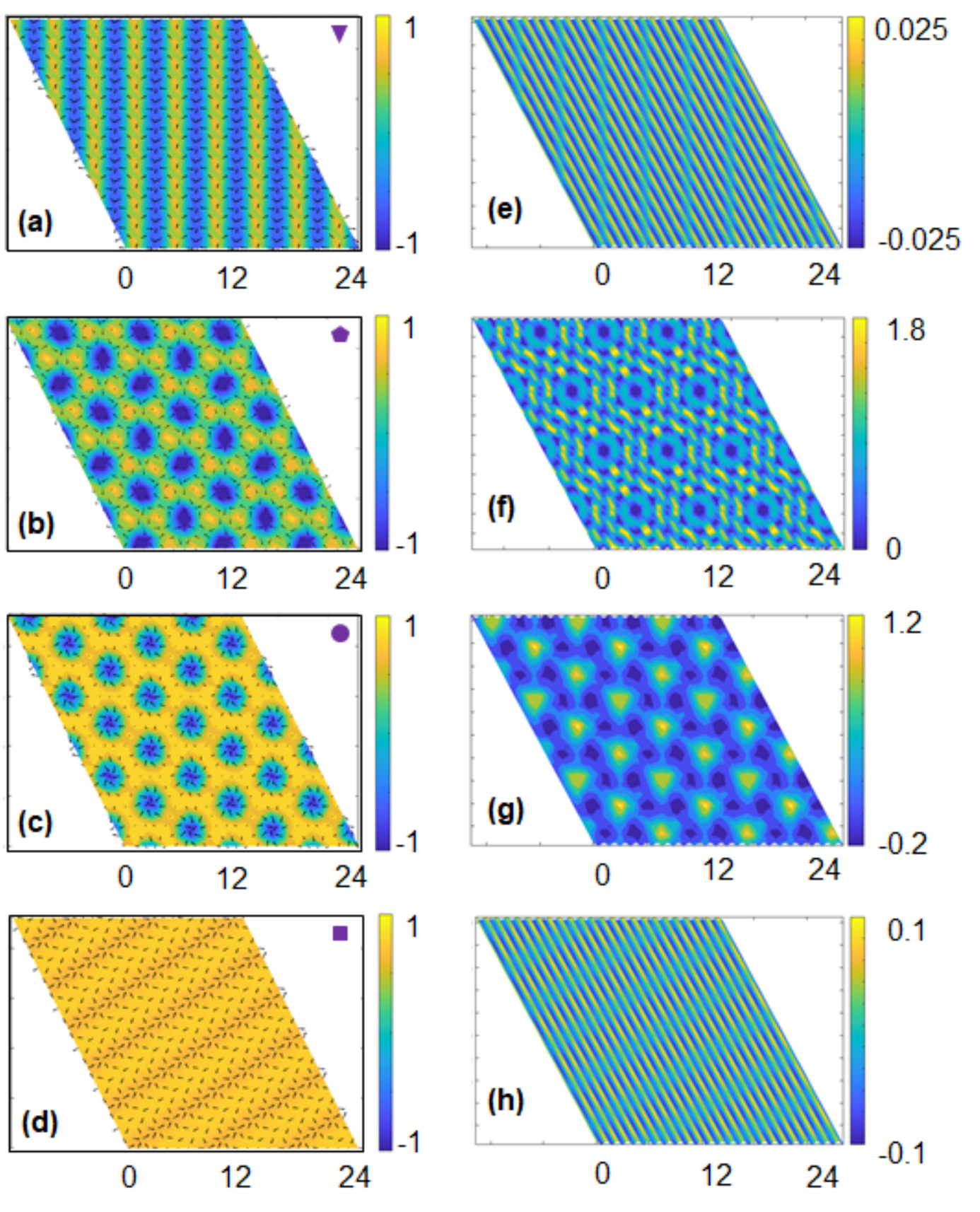}
\caption{Spin textures (a-d) and the corresponding topological charge densities (e-h) of NiIBr for magnetic fields $B_z$ = 1.2,2.1,7.3 and 9.8 meV (same as the ones in the main text).}
\label{tc_NiIBr}
\end{figure}

\begin{figure}[H]
\renewcommand{\figurename}{Figure}
\renewcommand{\thefigure}{S\arabic{figure}}
\centering
\includegraphics[width=8.4cm]{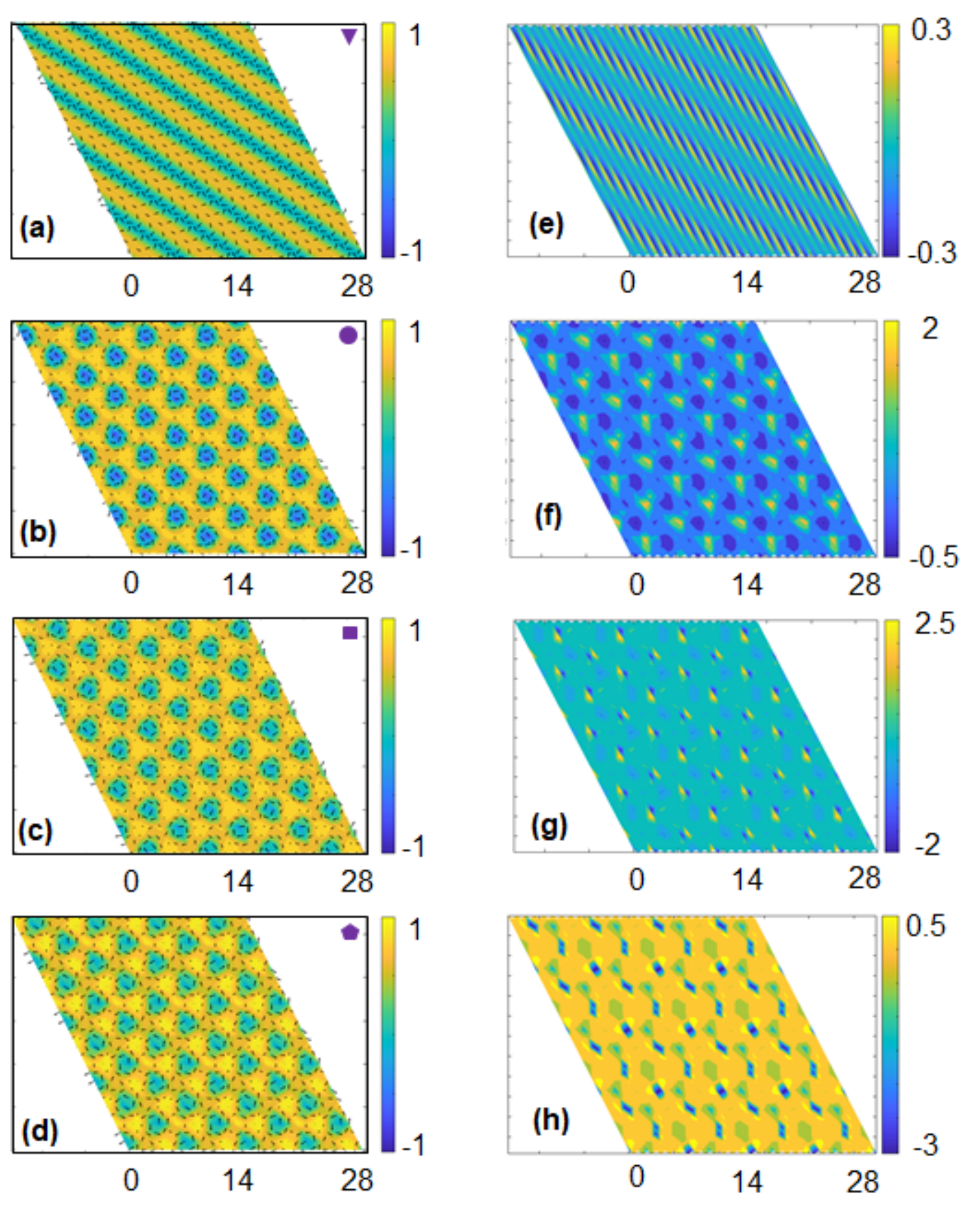}
\caption{Spin textures (a-d) and the corresponding topological charge densities (e-h) of NiICl for magnetic fields $B_z$ = 7.7,9.8,11.2 and 10.5 meV (same as the ones in the main text).}
\label{tc_NiICl}
\end{figure}

\begin{figure}[H]
\renewcommand{\figurename}{Figure}
\renewcommand{\thefigure}{S\arabic{figure}}
\centering
\includegraphics[width=8.4cm]{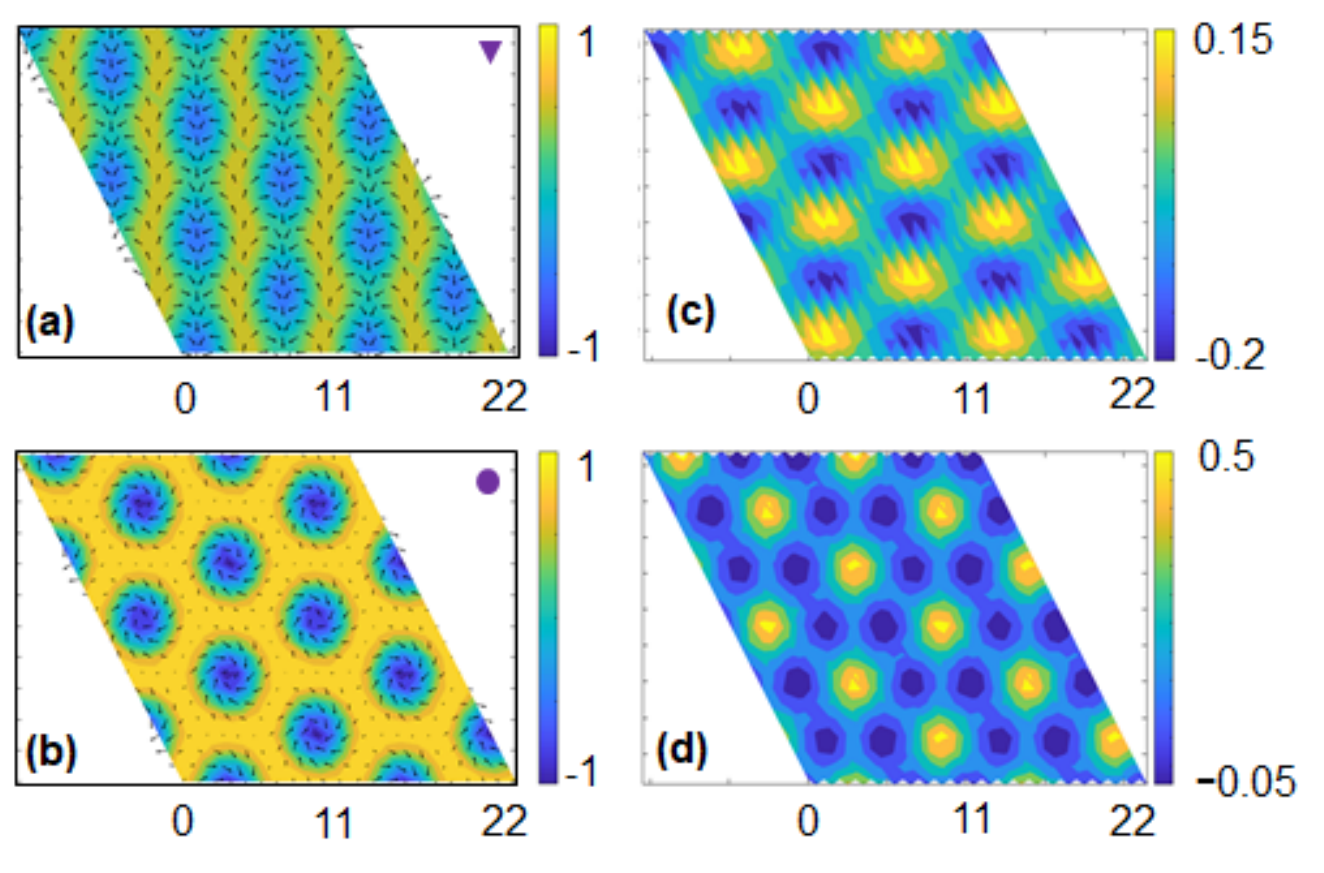}
\caption{Spin textures (a,b) and the corresponding topological charge densities (c,d) of NiClBr for magnetic fields $B_z$ = 0.5 and 1.3 meV (same as the ones in the main text).}
\label{tc_NiClBr}
\end{figure}

\bibliography{supporting}